\documentclass[preprint,11pt]{elsarticle}
\usepackage[toc,page]{appendix}
\usepackage{amsmath,amsthm,amsfonts,mathtools,bm}
\usepackage{multirow}
\usepackage{tikz,pgfplots}
\usetikzlibrary{patterns}
\definecolor{BurntOrange}{rgb}{0.8, 0.33, 0.0}
\usepackage{subcaption}

\pgfplotsset{compat=newest}
\usepackage{float,indentfirst,algorithm}
\usepackage{pgfplots}
\usepackage{booktabs}
\usepackage[margin=1in]{geometry}
\usepackage[frozencache =true,cachedir=.,outputdir=.]{minted}

\theoremstyle{plain}

\usepackage[capitalise]{cleveref}
\usepackage{appendix}
\usepackage{multicol}
\usepackage{nomencl}
\usepackage{ifthen}
\usepgfplotslibrary{groupplots}

\usepackage{todonotes}
\usepackage{url}

\newcommand{\bx}{\mathbf{x}}
\newcommand{\by}{\mathbf{y}}
\newcommand{\bu}{\mathbf{u}}

\newcommand{\RR}[0]{\mathbb{R}}

\makenomenclature
\usepackage{etoolbox}
\renewcommand\nomgroup[1]{%
  \item[\bfseries
  \ifstrequal{#1}{A}{Symbols}{%
  \ifstrequal{#1}{B}{Superscripts}{%
  \ifstrequal{#1}{C}{Subscripts}{
  \ifstrequal{#1}{D}{Operators}{}}}}%
]}

\makeatletter
\def\ps@pprintTitle{%
   \let\@oddhead\@empty
   \let\@evenhead\@empty
   \def\@oddfoot{\reset@font\hfil\thepage\hfil}
   \let\@evenfoot\@oddfoot
}
\makeatother

\begin{document}

\begin{abstract}
We propose a framework for training neural networks that are coupled with partial differential equations (PDEs) in a parallel computing environment. Unlike most distributed computing frameworks for deep neural networks, our focus is to parallelize both numerical solvers and deep neural networks in forward and adjoint computations. Our parallel computing model views data communication as a node in the computational graph for numerical simulations. The advantage of our model is that data communication and computing are cleanly separated and thus provide better flexibility, modularity, and testability. We demonstrate using various large-scale problems that we can achieve substantial acceleration by using parallel solvers for PDEs in training deep neural networks that are coupled with PDEs.
\end{abstract}


\begin{frontmatter}

\title{Distributed Machine Learning for Computational Engineering using MPI}
\author[rvt1]{Kailai~Xu}
\ead{kailaix@stanford.edu}

\author[rvt2]{Weiqiang Zhu}
\ead{zhuwq@stanford.edu}

\author[rvt1,rvt3]{Eric~Darve}
\ead{darve@stanford.edu}

\address[rvt1]{Institute for Computational and Mathematical Engineering,
               Stanford University, Stanford, CA, 94305}
\address[rvt3]{Mechanical Engineering, Stanford University, Stanford, CA, 94305}
\address[rvt2]{Department of Geophysics, Stanford University, Stanford, CA, 94305}

\end{frontmatter}

\section{Introduction}

Inverse modeling~\cite{vogel2002computational,kaipio2006statistical}, which aims at identifying parameters or functions in a physical model from observations, is one key technique to solving data-driven problems in many applications. With the advent of deep learning techniques \cite{goodfellow2016deep}, we are now able to build neural-network-based data-driven models to solve inverse problems involving more complex physical phenomena. Specifically, by substituting unknown components within a physical system described by partial differential equations (PDEs) with deep neural networks (DNNs) \cite{rackauckas2020universal,xu2020learning,goswami2020transfer,raissi2019physics,meng2020ppinn} , we can leverage the expressive power of DNNs while satisfying the physics to the largest extent. 

Gradient-based optimization algorithms \cite{boyd2004convex,luenberger1984linear} are usually applied to train DNNs that are coupled with PDEs. A loss function, which measures the discrepancy between true and hypothetical observations, is minimized by iteratively updating DNN weights and biases. The gradients can be computed using automatic differentiation \cite{paszke2017automatic,baydin2017automatic,abadi2016tensorflow} and adjoint state methods \cite{plessix2006review}. In our previous work \cite{xu2020physics,huang2020learning,xu2020learning} , we have developed a general framework, ADCME\footnote{\url{https://github.com/kailaix/ADCME.jl}}, that expresses both DNNs and numerical PDE solvers (e.g., finite element methods) as computational graphs. Therefore, we can calculate the gradients automatically and with machine accuracy using reverse-mode automatic differentiation (AD). In ADCME, the automatic differentiation operates on a higher level of abstractions, such as tensor operations and matrix solvers, or even PDE solvers,  instead of elementary ones (e.g., arithmetic operations) considered in general purpose AD software packages. This coarser granularity allows us to apply tensor-level performance optimizations and implement algorithms with a more intuitive and mathematical interface for scientific computing applications. 

\begin{figure}[htpb]
	\centering 
	\includegraphics[width=0.6\textwidth]{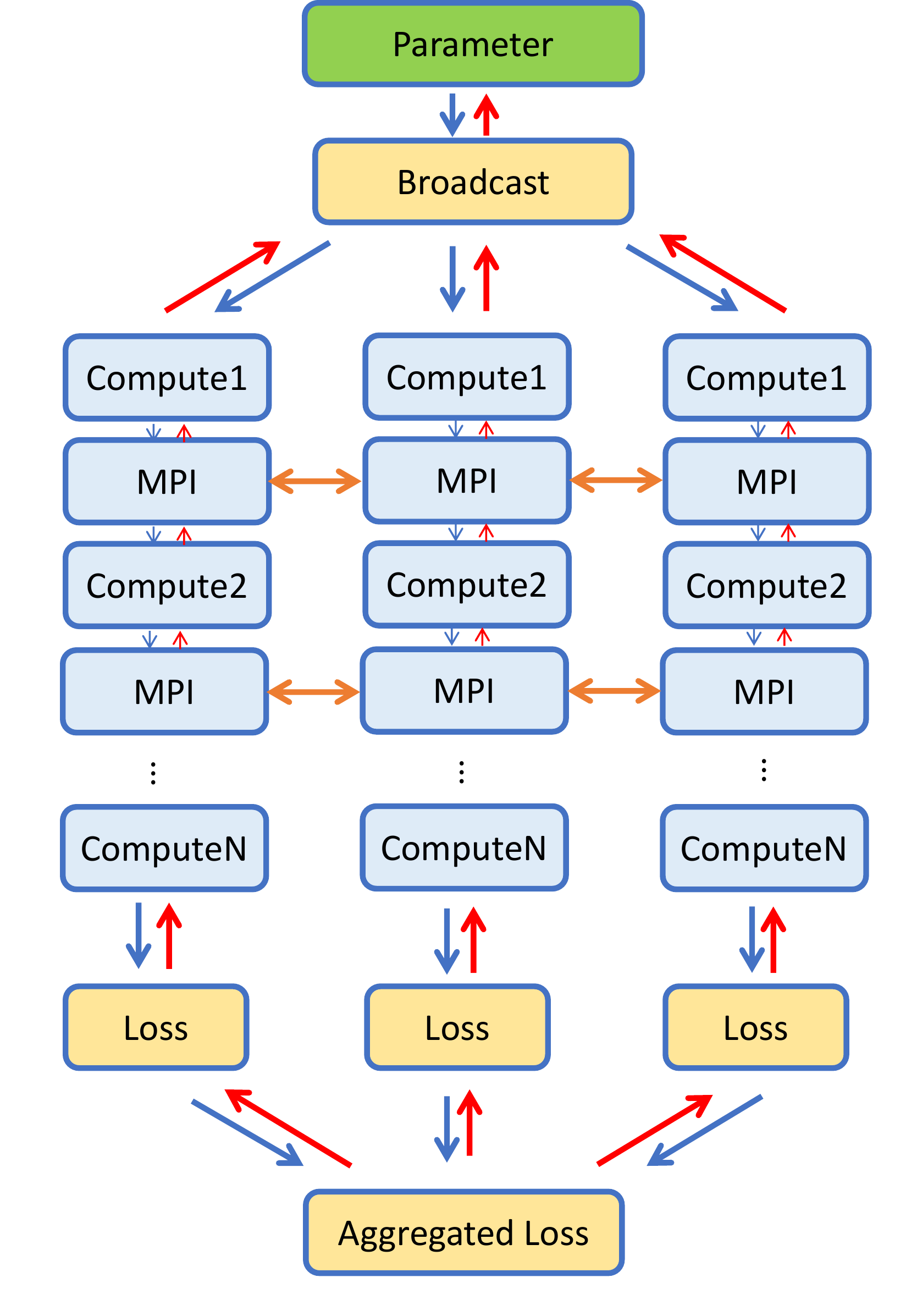}
	\caption{Parallel computational model for coupling DNNs and PDEs. There are two important data communication components: the parameters of DNNs are broadcasted to each MPI processor; within each MPI processor, there is data communication between adjacent patches in a computational grid.}
	\label{fig:ourmodel}
\end{figure}

However, as the problem size grows, the memory consumption becomes prohibitive because reverse-mode AD requires saving all intermediate results. For example, a direct AD implementation of a 2D elastic equation double-precision solver with a mesh size $1000\times 1000$ and $2000$ steps requires at least 193 gigabytes memory \footnote{We need to save around 13 vectors (wavefields, stress tensors, and auxiliary fields) of size $1000\times 1000\times 2000$.} \cite{zhu2020general}, which is impractical for a typical CPU. Parallel computing using MPI \cite{gabriel2004open,gropp1999using} is the de-facto standard for solving such a large-scale problem on modern distributed memory high performance computing (HPC) architectures. However, integrating a parallel computing capability into a reverse-mode automatic differentiation framework is challenging \cite{wang3high,utke2009toward,cheng2006duality}. We need to back-propagate gradients through DNNs, numerical solvers, and control flows of parallel communications (\Cref{fig:ourmodel}). These difficulties are magnified by the fact that we need the flexibility for hybrid threaded MPI capabilities (e.g., each MPI processor runs a multi-threading task and uses multiple cores on the same CPU) and therefore special efforts are needed to topological design of computational graph to avoid deadlocks.

\begin{figure}[htpb]
	\centering 
	\includegraphics[width=0.55\textwidth]{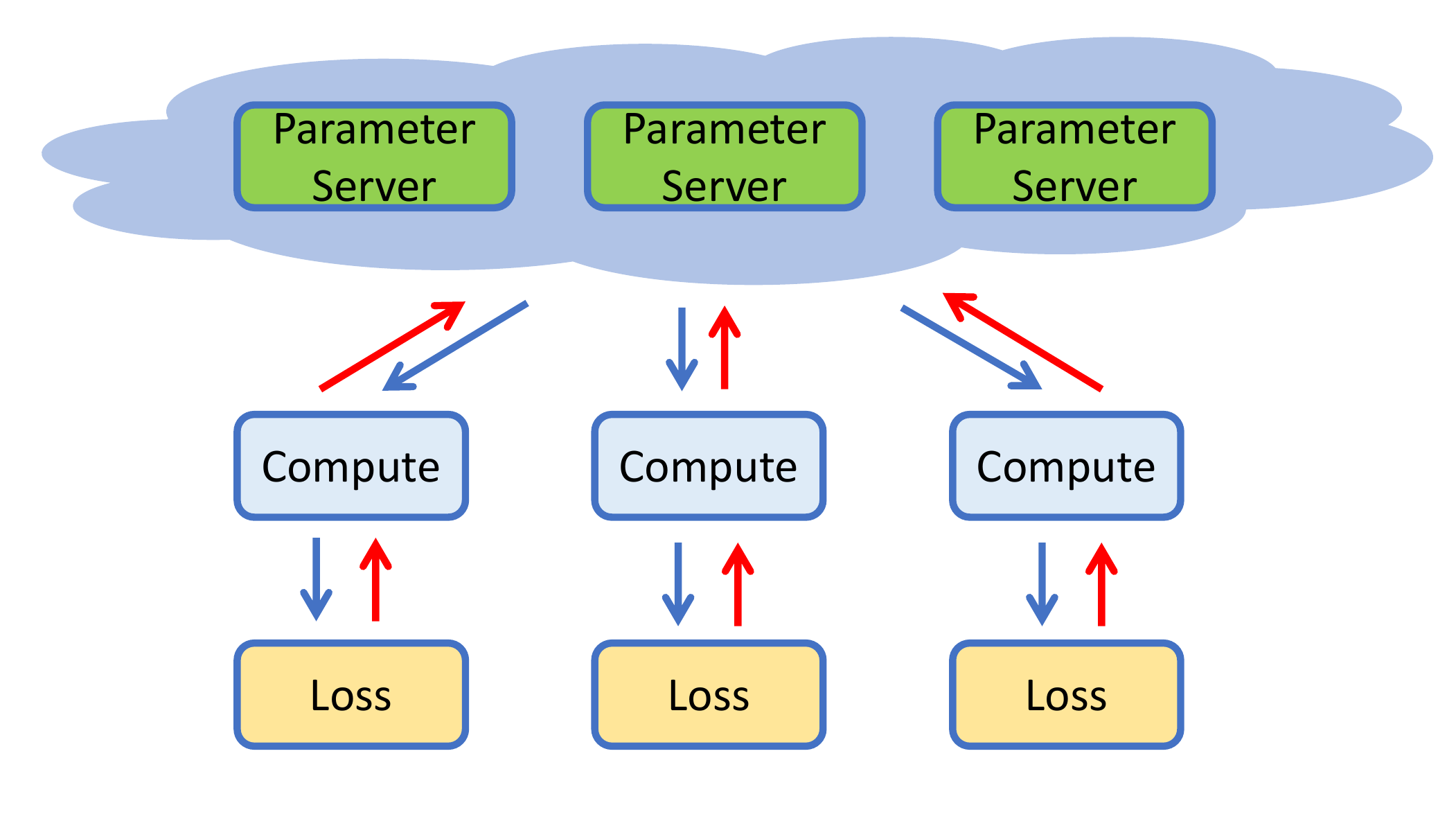}~
	\includegraphics[width=0.42\textwidth]{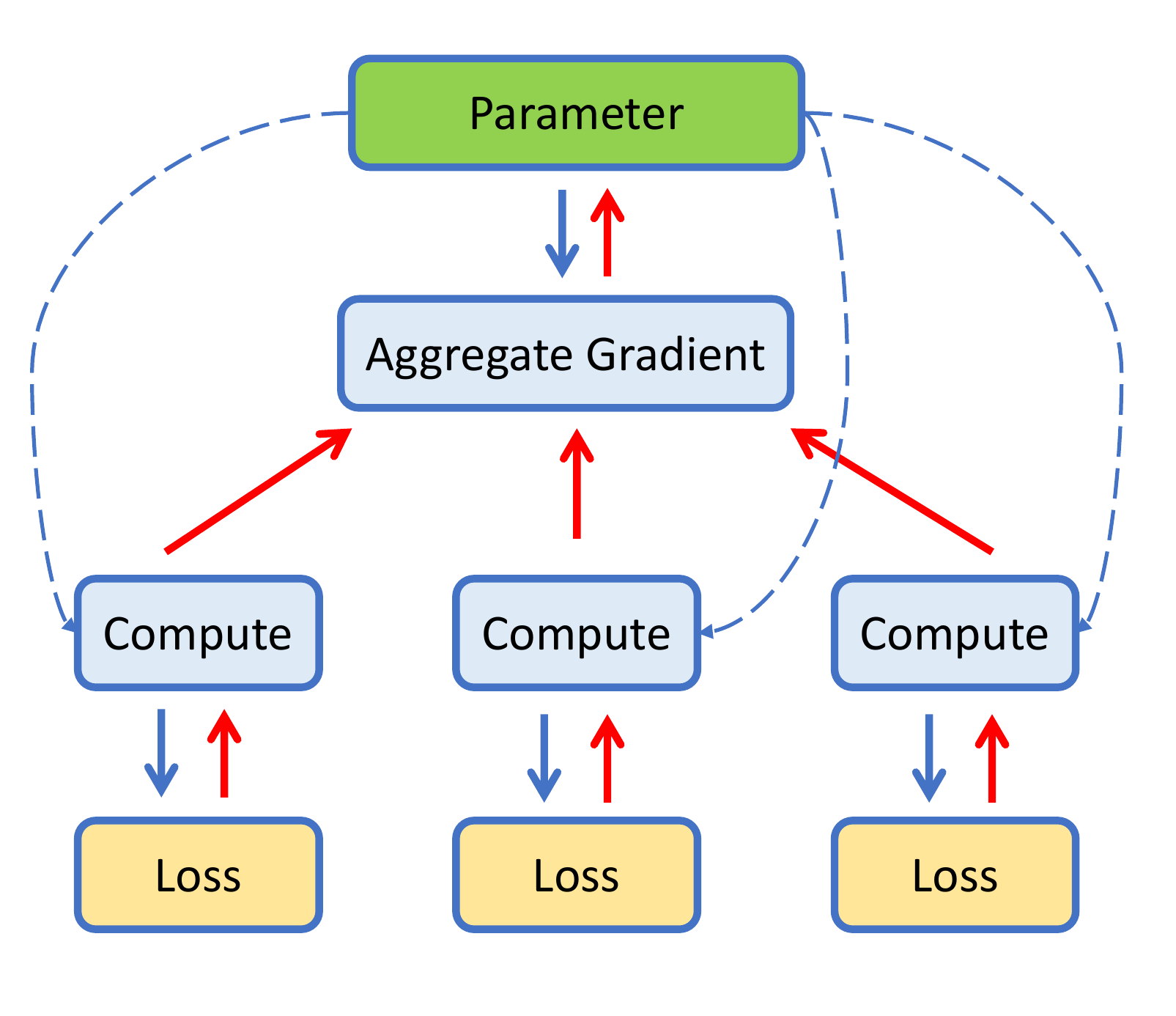}
	\caption{Left: parallelism via parameter servers. Each processor requests and updates states from the parameter servers. Parameter servers are responsible for maintaining the global states by communication within parameter servers themselves. Right: this paralell model can be viewed as a simplified version of parameter servers, where a master node collects all gradient information from other worker nodes. }
	\label{fig:parallel}
\end{figure}

Currently, distributed computing for reverse-mode AD have been intensively studied by two research communities: parallelizing numerical solvers \cite{griebel1999parallel,notay1995efficient,douglas2003tutorial} in scientific computing, and parallelized training (e.g., DNNs) \cite{bekkerman2011scaling,jordan2015machine} in machine learning. In the scientific computing community,  a computational grid is usually split into multiple patches and each MPI processor owns one patch. This allows to distribute degrees of freedoms onto different machines and thus relaxes the memory demands. Examples include \texttt{libadjoint} \cite{farrell2013automated,mitusch2019dolfin} and \texttt{DAfoam} \cite{he2020dafoam}. However, these frameworks are domain specific and are tightly integrated into \texttt{dolfin} \cite{dupont2003fenics} and \texttt{OpenFOAM} \cite{openfoam}, respectively. This integration makes them difficult to adapt for general purpose data-driven inverse modeling, especially when deep neural networks are present and coupled with PDEs. In the machine learning community, one popular model is data-parallelism distributed computing \cite{bennun2018demystifying}, where each machine has full information of a complete DNN but only a portion of training data \cite{verbraeken2020survey}. These DNNs share the same parameters, through a synchronization mechanism such as parameter servers (\Cref{fig:parallel}) \cite{li2014scaling}. In these cases, the computations are mostly independent and there is no communication among different processors within each computational graphs. There are also the so-called model-parallelism distributed computing, in which DNNs themselves are split onto different machines  \cite{chen2018efficient}. However, these methods eventually lead to asynchronous training, where the correctness of gradients or the convergence is not guaranteed. Note that the distributed computing models from both two communities are quite different and are designed to meet their own needs. To couple DNNs and PDEs, where both synchronization of DNN parameters and parallel PDE solvers are present, a new distributed computing model that combines their features is desirable.

\begin{figure}[htpb]
	\centering 
	\includegraphics[width=0.48\textwidth]{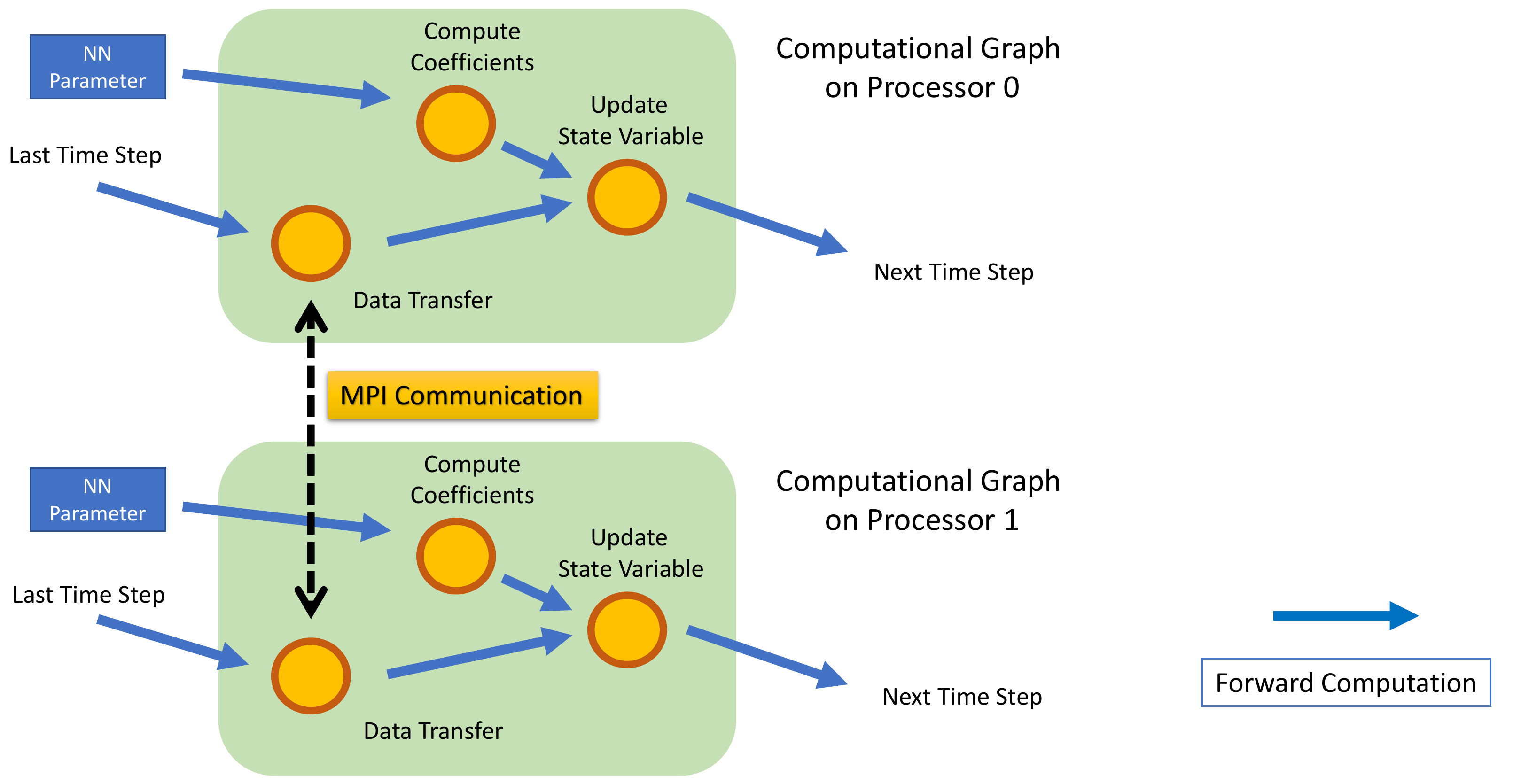}~
		\includegraphics[width=0.48\textwidth]{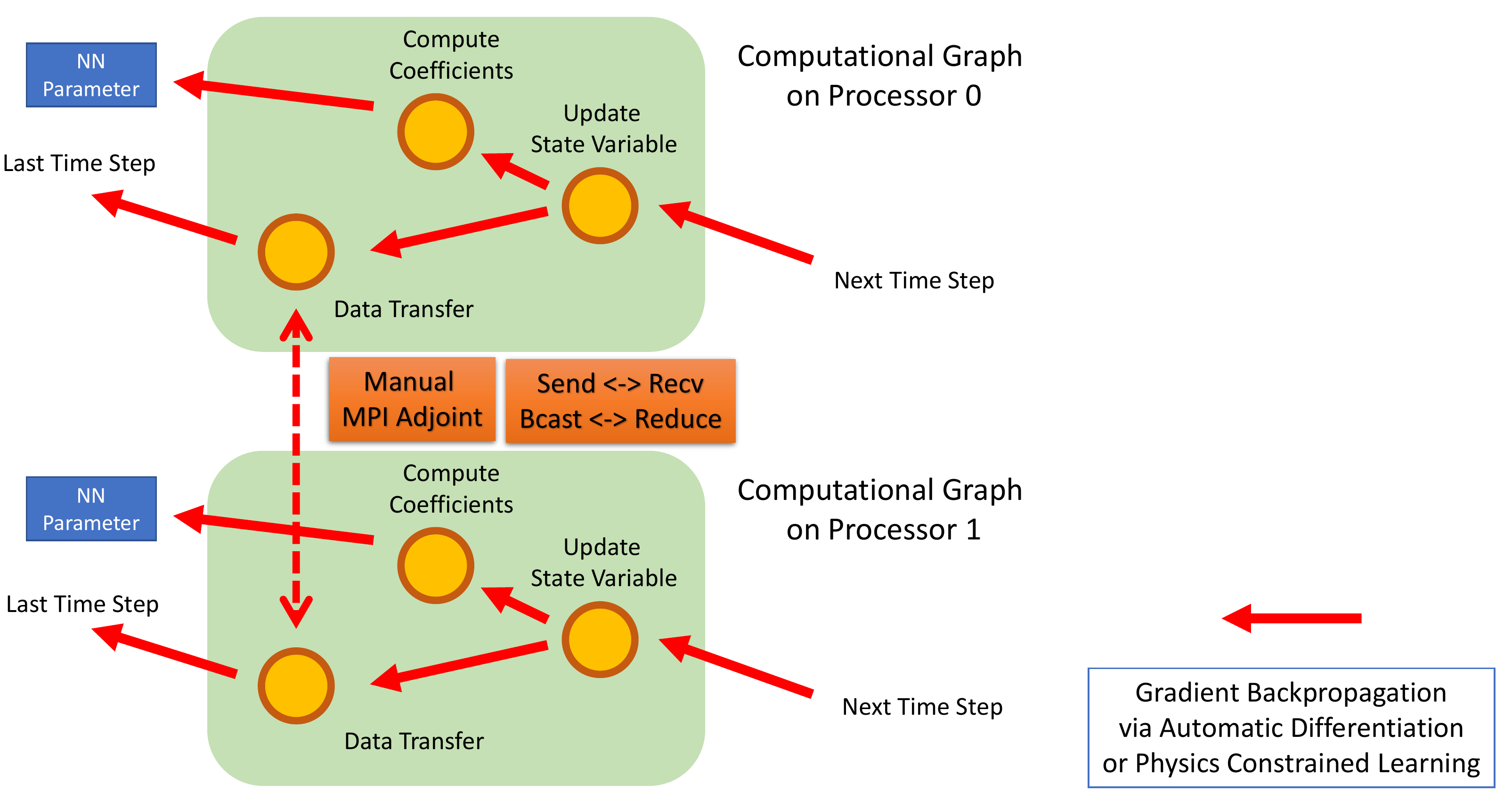}
		\caption{Paradigm for distributed machine learning for computational engineering. The data communication operations are treated as nodes in the computational graph. They are cleanly separated from other computing operators and therefore we can reuse serial implementations of the numerical PDE solver.}
		\label{fig:paradigm}
\end{figure}

In this work, we propose a novel framework for developing MPI-based distributed reverse-mode AD algorithms for coupling DNNs and PDEs. We consider treating data communication within both PDE solvers and DNN updates as nodes (operators) in the computational graph (\Cref{fig:paradigm}). These operators are non-intrusive to the other operators in the sense that existing implementation of the latter operators also works for parallel execution. The clean separation of data communication and computing offers the flexibility of reusing the existing ADCME framework and having the same implementation for both serial and parallel computing environments. 

We use the high level language Julia \cite{bezanson2017julia} to specify the dependency of data communication operators and other computation operators. These data communication operators can be either built-in ADCME operators (e.g., \texttt{mpi\_bcast}, which broadcasts a tensor to all workers), or provided by users in the form of custom operators. All computation and data communication are delegated to C++ or CUDA kernels at runtime, and the correctness of parallel programs is ensured by the topology of the computational graph. The MPI capability of ADCME relieves us from reasoning about the order of MPI calls, e.g., carefully ordering sends and receives to avoid deadlocks, and parallel communication rules for AD, e.g., forward sends become backward receives. Therefore, ADCME allows us to manage complex numerical simulations and accelerate development cycles. The separation of data communication and computation operators also allows us to test, analyze, and optimize them individually.

The aim of this paper is to describe the mechanism underlying ADCME MPI-based distributed computing capability. We also apply the tool to compute DNN gradients in a coupled system of DNNs and PDEs, such as Poisson's equations and wave equations. We also present some important operators and data structures for scientific computing. For example, we use Hypre \cite{falgout2002hypre} as the distributed sparse linear solver backend and use CSR formats (\texttt{mpi\_SparseTensor}) for storing a sparse matrix on each machine. In gradient back-propagation, the transpose of the original matrix is usually needed and such functionalities are missing in Hypre as of September 2020. The transposition is implemented as a separate operator in ADCME. The numerical examples demonstrate that ADCME can scale up to thousands of cores and solve large-scale scientific machine learning problems.

\section{Distributed Scientific Machine Learning}

\subsection{Machine Learning for Computational Engineering}

As an example, we consider the Poisson's equation with a spatially varying diffusivity coefficient  $\kappa(\bx)$:
\begin{equation}\label{equ:poisson}
	\begin{aligned}
		\nabla \cdot (\kappa(\bx) \nabla u(\bx)) & = f(\bx) &\bx \in \Omega \\ 
		u(\bx) &= 0 &\bx \in \partial \Omega
	\end{aligned}
\end{equation}
Here $f(\bx)$ is a given source function, and $\Omega$ is a domain in $\RR^d$. Assume that we can observe the state variable $u(\bx)$ at some locations $\{\bx_i\}_{i\in \mathcal{I}}$, denoted as $\{u_i\}_{i\in \mathcal{I}}$, where $\mathcal{I}$ is the index set. We want to estimate $\kappa(\bx)$, whose form is in unspecified, from the observations. 

This inverse problem can be formulated as an optimization problem
\begin{equation}\label{equ:min}
	\min_{\kappa\in\mathcal{K}} \sum_{i\in \mathcal{I}} \left( u(\bx_i) - u_i \right)^2
\end{equation}
Here $u(\bx_i)$ are the solutions to \Cref{equ:poisson} for a given $\kappa(\bx)$, and $\mathcal{K}$ is an appropriate function space for the unknown $\kappa$. 

However, \Cref{equ:min} is typically an infinitely dimensional optimization problem if $\mathcal{K}$ cannot be parametrized by parameters with a finite dimension. To solve it numerically, we can approximate $\kappa$ with a deep neural network 
$$\kappa(\bx)\approx \mathcal{N}_\theta (\bx)$$
where $\theta$ is the neural network weights and biases. This approach has been shown to be very effective, especially when $\kappa$ is a high dimensional mapping or lacks smoothness \cite{xu2020calibrating}, compared to traditional approaches such as piecewise linear functions. Additionally, DNN representation also provides regularization thanks to its spatial correlation (e.g., if $\bx_1$ and $\bx_2$ are close, $\kappa(\bx_1)$ and $\kappa(\bx_2)$ are also close) and appropriate initialization. 

Therefore, the optimization problem is reduced to a finite dimensional optimization problem, where the optimization variable is the DNN parameters:
 \begin{equation}\label{equ:optimization}
 	\begin{aligned}
 		\min_{\theta} &\; J(\theta):=\sum_{i\in \mathcal{I}} \left( u(\bx_i) - u_i \right)^2\qquad (u(\bx_i)\text{ is an indirect function of }\theta) \\
 	\text{s.t.} &\;	\nabla \cdot (\mathcal{N}_\theta(\bx) \nabla u(\bx)) = f(\bx) \qquad \bx \in \Omega \\ 
 	 &\;	u(\bx) = 0 \qquad \bx \in \partial \Omega
 	\end{aligned}
 \end{equation}

The PDE constraint in \Cref{equ:optimization} can be numerically solved using finite difference methods (or finite element methods), which results in a linear system 
\begin{equation}\label{equ:auf}
\mathbf{A} \mathbf{u} = \mathbf{f}
\end{equation}

Here $\bu\in \RR^n$, $\mathbf{f}\in \RR^n$, $\mathbf{A}\in\RR^{n\times n}$, and $n$ is the degrees of freedoms. The reconstructed numerical solution $u(\bx)$ depends on $\theta$, although in general, we can not find an explicit expression for $u(\bx)$ with respect to $\theta$. 

To solve the optimization problem \Cref{equ:optimization}, we will calculate the gradients $\nabla_\theta J(\theta)$. In the context of reverse-mode automatic differentiation, this calculation requires us to back-propagate gradients through both the numerical solver and the neural network. In our previous work, we proposed a general framework for expressing both numerical simulations and DNNs as computational graphs, and thus unified the implementation of reverse-mode AD for deep neural networks and numerical solvers. This technique enables us to easily extract gradients $\nabla_\theta J(\theta)$ by reverse-mode automatic differentiation.

\subsection{Parallel Computing}

One challenge associated with reverse-mode AD is that all intermediate variables must be saved for gradient back-propagation. This requires additional memory compared to  forward computation. Large-scale problems bring another difficulty to memory costs: we may not be able to solve even the forward problem, because it is challenging to save a large-sized discrete solution $\mathbf{u}$ on a single CPU. 

One solution to ensure performance and scalability of numerical solvers is MPI. When we use MPI to solve \Cref{equ:optimization}, the grid is split into multiple patches, and each machine (an MPI processor) owns one patch. We can rearrange the indices so that each MPI processor owns a stripe of rows in the matrix $\mathbf{A}$ and a segment of $\mathbf{f}$ with continuous indices. We resort to a distributed solver (e.g., from Hypre \cite{falgout2002hypre}) to solve the linear system \Cref{equ:auf}. After solving the linear system, each patch owns a segment of $\bu$, and a local loss function can be computed. The local loss functions are then aggregated to calculate the global loss function. 

However, when we compute the gradients, reverse-mode automatic differentiation with MPI is challenging because reasoning about reverse rules is difficult while avoiding deadlocks, ensuring correctness, and improving performance. There are many researches that focus on adjoint MPI \cite{he2020dafoam,towara2015mpi}. Many of them are tightly coupled with a specific application, and/or are not developed in the context of DNNs and PDEs. 

Note that there are also active research in the deep learning community for distributed computing of DNNs. However, the parallel models are usually data-parallelism and do not consider communication within layers. For model parallelism in DNNs \cite{hewett2020linear}, usually DNNs are split into several parts and an asynchronous training method is used. These models are not appropriate for the inverse modeling problems that we are considering because communication does not happen within computational graphs. Instead, in our parallel model, the data communication pattern is more complicated: the data communication may occur multiple times within the computational graph, e.g., while performing time-stepping schemes or accumulating the loss function, etc. This difference calls for a new parallel model for both forward computation and gradient back-propagation.

\section{Distributed Computing Design in ADCME}

\subsection{Basic MPI Operators}

ADCME provides a subset of the MPI routines to facilitate implementing computational graphs with data communication capabilities. Most of the routines have gradient back-propagation capabilities. For example, the \texttt{mpi\_gather} function corresponds to \texttt{MPI\_Gather}, except that \texttt{mpi\_gather} is differentiable. Internally, this API is implemented by observing that the scatter operation is the reverse of the gather operation in an MPI program. We show an excerpt from the \texttt{mpi\_gather} implementation in ADCME:
\begin{minted}{cpp}
void MPIGather_forward(double *out, const double *a, int m, int root){
     MPI_Comm comm = MPI_COMM_WORLD;
     MPI_Gather( a , m , MPI_DOUBLE , out , m , MPI_DOUBLE , root , comm);
}

void MPIGather_backward(
double *grad_a, const double *grad_out,
const double *out, const double *a, int m, int root
){
     MPI_Comm comm = MPI_COMM_WORLD;
     MPI_Scatter( grad_out , m , MPI_DOUBLE , grad_a ,  m , MPI_DOUBLE , root , comm);
}
\end{minted} 
Here \texttt{MPIGather\_forward} implements the forward computation, and broadcast \texttt{a} from the root to all worker nodes buffers \texttt{out}. \texttt{MPIGather\_backward} is the corresponding gradient back-propagation implementation. Having received \texttt{grad\_out} from the downstream of the computational graph, this \texttt{mpi\_gather} operator back-propagates the gradient to \texttt{grad\_a} by reversing \texttt{MPI\_Gather}. \Cref{tab:mpi} presents parts of the built-in MPI operators in ADCME.

\begin{table}[htpb]
	\centering
	\caption{MPI routines in ADCME. Many operations can perform both forward computation and gradient back-propagation. ``YES'' indicate that the operator supports gradient back-propagation.}
	\label{tab:mpi}
	\begin{tabular}{@{}lll@{}}
		\toprule
		API              & Description                                                     & Gradient \\ \midrule
		mpi\_init        & Initialize an MPI session                                       & NO       \\
		mpi\_finalize    & Finalize an MPI session                                         & NO       \\
		mpi\_initialized & A boolean indicating whether an MPI session is initialized      & NO       \\
		mpi\_finalized   & A boolean indicating whether an MPI session is finalized        & NO       \\
		mpi\_rank        & Current processor rank                                          & NO       \\
		mpi\_size        & Total number of processors                                      & NO       \\
		mpi\_sync!       & Broadcast a Julia array on all processors                            & NO       \\
		mpi\_sum         & Sum a tensor on the designated processor                        & YES      \\
		mpi\_bcast       & Broadcast a tensor from thedesignated processor                 & YES      \\
		mpi\_recv        & Receive a tensor from a given processor                         & YES      \\
		mpi\_send        & Send a tensor from a given processor                            & YES      \\
		mpi\_sendrecv    & Send a tensor from a given processor to another given processor & YES      \\
		mpi\_gather      & Concat tensors from all processors on the root processor        & YES      \\ \bottomrule
	\end{tabular}
\end{table}

\begin{figure}[htpb]
	\centering 
	\includegraphics[width=0.6\textwidth]{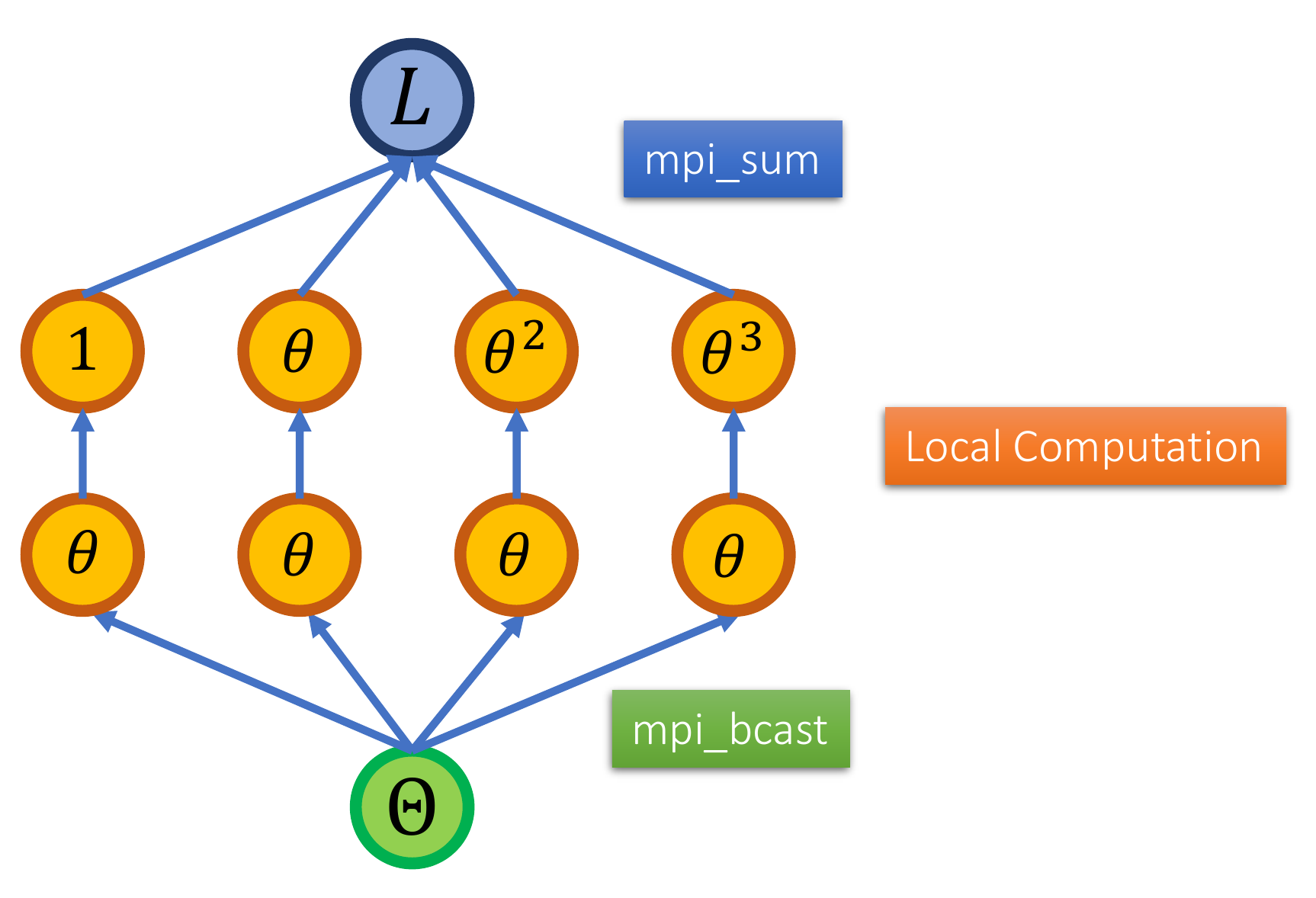}
	\caption{An illustration for computing \Cref{equ:simple}. $\theta$ is broadcasted onto 4 MPI processors. Local computing is then performed on each processor. Finally, the function values are aggregated via \texttt{mpi\_sum}.}
	\label{fig:simple}
\end{figure}

As an example on how to use these primitive operators, consider computing
\begin{equation}\label{equ:simple}
L(\theta) = 1 + \theta + \theta^2 + \theta^3
\end{equation}
The parameter $\theta$ is first broadcasted onto 4 processors using \texttt{mpi\_bcast}. Processor $i$ computed $\theta^i$ 
locally. Then the results are summed on the root processor using \texttt{mpi\_sum}. An ADCME program is as follows (\Cref{fig:simple})
\begin{minted}{julia}
using ADCME
mpi_init() # initialize MPI 
theta0 = placeholder(1.0)
theta = mpi_bcast(theta0)
l = theta^mpi_rank()
L = mpi_sum(l)
# initialize a Session
sess = Session(); init(sess)
L_value = run(sess, L) # only the value on rank 0 is valid
mpi_finalize() # finalize MPI 
\end{minted}
Additionally, we can use the following one-liner to extract gradients
\begin{minted}{julia}
grad = gradients(L, theta0)
\end{minted}
This simple line of code hides the details for back-propagating gradients through local operations as well as MPI routines (\texttt{mpi\_bcast}, \texttt{mpi\_sum}).

\subsection{Halo Exchange}

\begin{figure}[htpb]
	\centering 
	\includegraphics[width=0.3\textwidth]{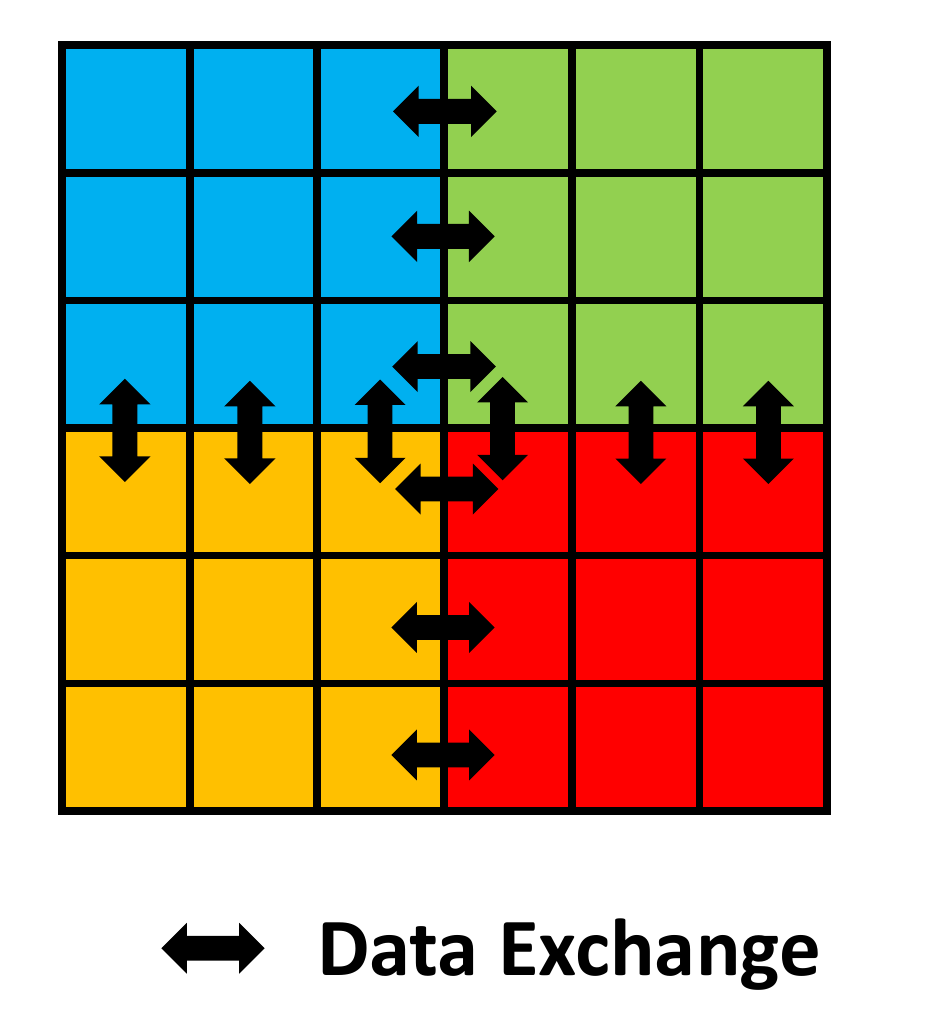}
	\caption{Domain decomposition and halo exchange. Two adjacent patches exchange boundary information.}
	\label{fig:halo}
\end{figure}

In many applications, we use domain decomposition to distribute the workload onto different processors. To reduce the data communication overheads, subdomains usually overlap at the boundaries and exchange boundary data with their adjacent subdomains (\Cref{fig:halo}). This communication pattern is referred to as halo exchange.  Halo exchange allows for communicating only a small portion of data compared to the data used for stencil computation within the subdomain. Halo exchange appears in many applications, and thus is implemented as a standalone operator in ADCME. 

\begin{figure}[htpb]
	\centering 
	\includegraphics[width=0.8\textwidth]{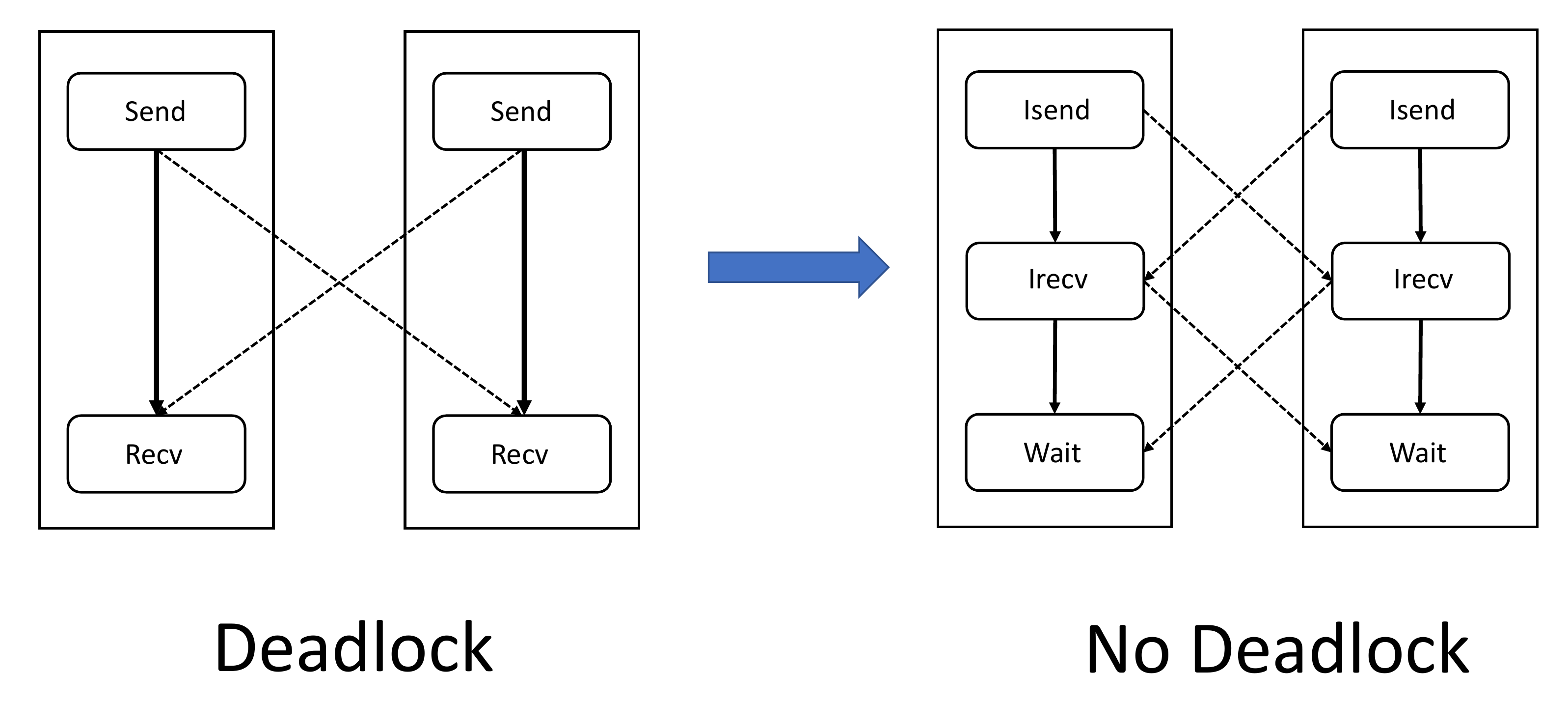}
	\caption{Avoid deadlocks using nonblocking sends and receives. Here \texttt{Isend} and \texttt{Irecv} are non-blocking sends and receives.}
	\label{fig:deadlock}
\end{figure}

The implementation of forward computation of halo exchange is straightforward. A direct application of blocking sends and receives will cause deadlocks. Instead, we use nonblocking sends and receives (\Cref{fig:deadlock}). For the reverse mode, the communication is also reversed for gradient back-propagation. The idea is that the forward sends correspond to reverse receives and the forward receives correspond to reverse sends. Additionally, the order should also be reversed. Once we work out the one-by-one substitution, the algorithm will work out correctly.  

\subsection{Distributed Sparse Matrices}

ADCME uses Hypre as the distributed linear algebra backend for MPI matrix solvers.  Each MPI processor owns a stripe of rows with continuous indices. The local matrix is stored in CSR (compressed sparse row) format and ADCME provides an interface \texttt{mpi\_SparseTensor} to hide the implementation details. 

For reverse-mode automatic differentiation, matrix transposition is an operator that are common in gradient back-propagation. For example, assume the forward computation is ($\bx$ is the input, $\by$ is the output, and $\mathbf{A}$ is a matrix)
\begin{equation}\label{equ:Axb}
	\by = \mathbf{A}\bx
\end{equation}
Given a loss function $L(\by)$, the gradient back-propagation calculates
$$\frac{\partial L(\by(\bx)))}{\partial \bx} = \frac{\partial L(\by)}{\partial \by} \frac{\partial \by(\bx)}{\partial \bx} = \frac{\partial L(\by)}{\partial \by} \mathbf{A}$$
Here $ \frac{\partial L(\by)}{\partial \by}$ is a row vector, and therefore 
$$\left(\frac{\partial L(\by(\bx)))}{\partial \bx}\right)^T = \mathbf{A}^T \left(\frac{\partial L(\by)}{\partial \by}\right)^T$$
requires a matrix vector multiplication, where the matrix is $\mathbf{A}^T$. 

Given that \Cref{equ:Axb} is ubiquitous in numerical PDE  schemes, a distributed implementation of parallel transposition is very important. As mentioned, ADCME distributes a continuous set of rows in a sparse matrix onto different MPI processors. The sub-matrices are stored in the CSR format. To transpose the sparse matrix in a parallel environment, we first split the matrices in each MPI processor into blocks and then use \texttt{MPI\_Isend}/\texttt{MPI\_Irecv} to exchange data. Note that this procedure consists of two phases: in the first phase, each block determines the number of nonzero entries to send, and sends the number to its corresponding receiver; in the second phase, each block actually sends all nonzero entries. Since the receiver already knows the number of entries expected to receive, it can prepare a buffer of an appropriate size. Finally, we transpose the matrices in place for each block. Using this method, we obtained a CSR representation of the transposed matrix (\Cref{fig:transpose}).
The gradient back-propagation is also implemented accordingly by reversing all the communication steps. 

\begin{figure}[htpb]
	\centering 
	\includegraphics[width=0.8\textwidth]{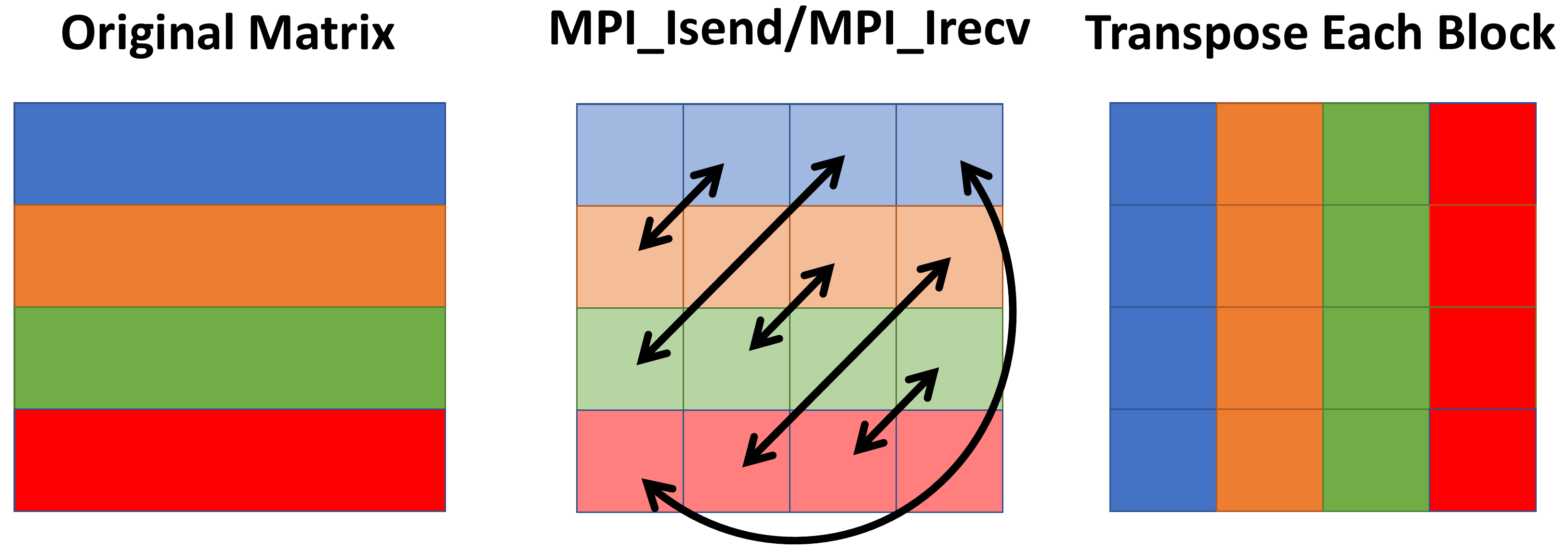}
	\caption{Transposition of a distributed sparse matrix. Each block exchange nonzero entries with the corresponding transposed block. The original and resulting matrices are both stored in CSR formats.}
	\label{fig:transpose}
\end{figure}

\subsection{Distributed Optimization}\label{sect:distopt}

The objective function of our problem can be written as a sum of local objective functions
$$\min_{\theta} L(\theta) = \sum_{i=1}^N f_i(\theta)$$
where $f_i$ is the local objective functions, $N$ is the number of processors.

Despite many existing distributed optimization algorithm, in this work we adopt a simple approach: aggregating gradients $\nabla_\theta f_i(\theta)$ and updating $\theta$ on the root processors. Note this is possible because in scientific computing applications, neural networks are not necessarily very large and therefore the dimension of $\theta$ is typically small. This  approach allows us to apply sophisticated gradient-based optimization problem easily with existing off-the-shelf optimizers, such as L-BFGS \cite{liu1989limited}, nonlinear conjugate gradient method \cite{hager2006survey}, and ADAM optimizer \cite{kingma2014adam}.

\begin{figure}[htpb]
	\centering 
	\includegraphics[width=0.8\textwidth]{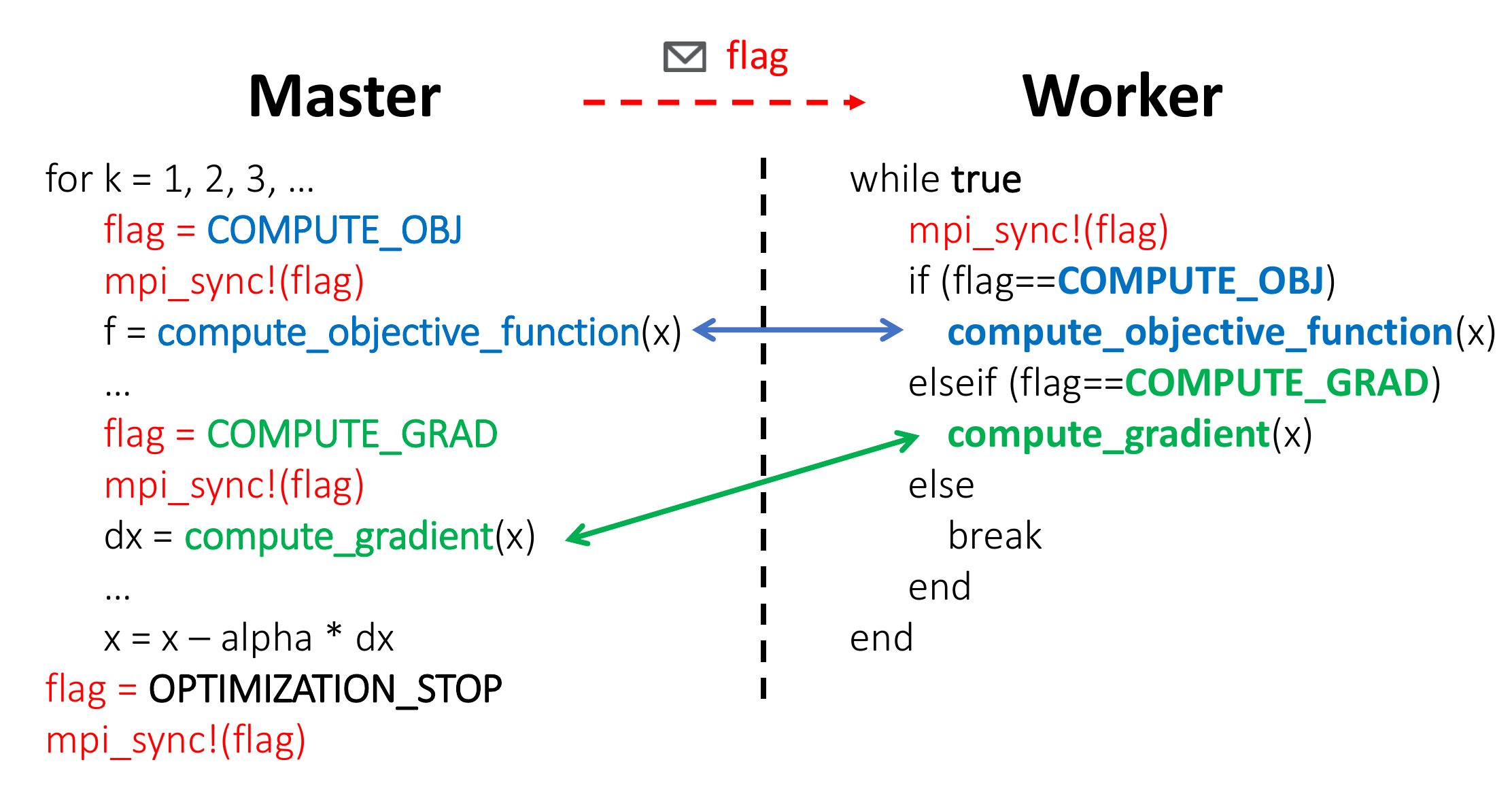}
	\caption{Refactor a serial optimizer to MPI-based optimizer in our framework.}
	\label{fig:optimization}
\end{figure}

\Cref{fig:optimization} shows how we can convert an existing optimizer to an MPI-enabled optimizer. The basic idea is to let the root processor notify worker processors whether to compute the loss function or the gradient. Then the root processor and workers will collaborate on executing the same routines and thus ensuring the correctness of collective MPI calls.

\subsection{Hybrid Programming}

Each MPI processor can communicate data between processes, which do not share memory. Within each process, ADCME also allows for multi-threaded parallelism with a shared-memory model. For example, we can use OpenMP to accelerate matrix vector production. We can also use a threadpool per process\footnote{In fact, the TensorFlow backend has two threadpools: one for inter-parallelism, i.e., independent operators can be executed concurrently, and another one for intra-parallelism, i.e., multi-threading within an individual operator.} to manage more complex and dynamic parallel tasks. However, the hybrid model brings challenges to communicate data using MPI. When we post MPI calls from different threads within the same process, we need to prevent data races and match the corresponding broadcast and collective operators. For example, without any guarantee on the ordering of concurrent MPI calls, we might incorrectly matched a send operator with a gather operator. 

\begin{figure}[htpb]
	\centering 
	\includegraphics[width=0.8\textwidth]{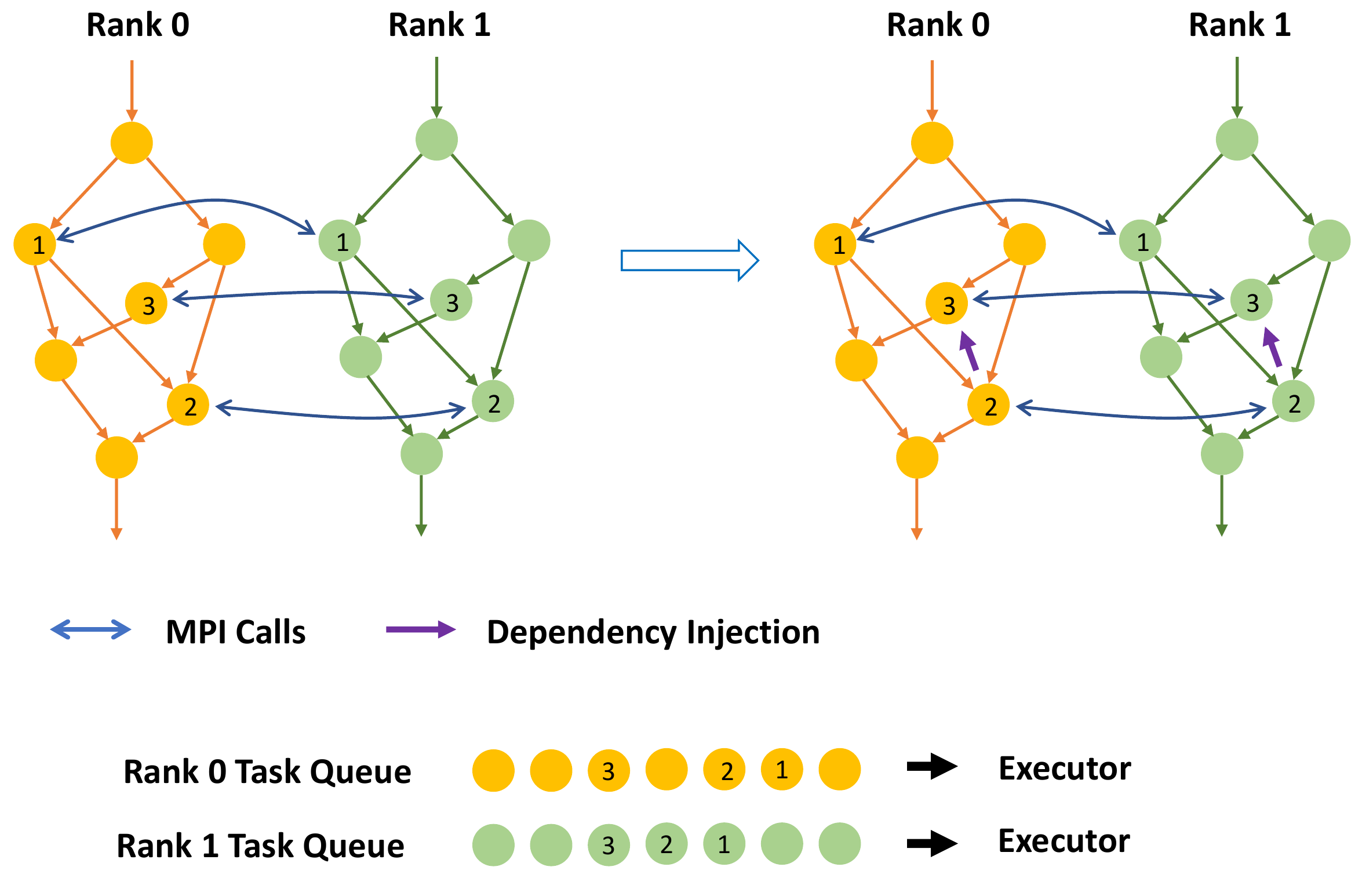}
	\caption{Ensuring the correctness of multi-threaded MPI programs using dependency injection. We make Operator 3 explicitly depend on Operator 2. The bottom shows the working queues for each MPI processor, the MPI operators are serialized in the same order.}
	\label{fig:hybrid}
\end{figure}

In ADCME, we adopt the dependency injection \cite{chiba2005aspect} technique: we explicitly serialize the MPI calls by adding ghost dependencies. For example, in the computational graph in \Cref{fig:hybrid}, originally, Operator 2 and Operator 3 are independent. In a concurrent computing environment, Rank 0 may execute Operator 2 first and then Operator 3, while Rank 1 executes Operator 3 first and then Operator 2. Then there is a mismatch of the MPI call (race condition): Operator 2 in Rank 0 coacts with Operator 3 in Rank 1, and Operator 3 in Rank 0 coacts with Operator 2 in Rank 1. 

To resolve the data race issue, we can explicitly make Operator 3 depend on Operator 2. In this way, we can ensure that \textbf{the MPI calls} Operator 1, 2, and 3 are executed in order. Note this technique sacrifices some concurrency (Operator 2 and Operator 3 cannot be executed concurrently), but the concurrency of most non-MPI operators is still preserved. 

\section{Benchmarks}

The purpose of this section is to present the distributed computing capability of ADCME via MPI. With the MPI operators, ADCME is well suited to parallel applications on clusters with very large numbers of cores. We benchmark individual operators as well as the gradient calculation as a whole. Particularly, we use two metrics for measuring the scaling of the implementation:
\begin{enumerate}
	\item \textbf{Weak scaling}, i.e., how the solution time varies with the number of processors for a fixed problem size per processor.
	\item \textbf{Strong scaling}, i.e., the speedup for a fixed problem size with respect to the number of processors, and is governed by Amdahl's law.
\end{enumerate}
For most operators, ADCME is just a wrapper of existing third-party parallel computing software (e.g., Hypre). However, for gradient back-propagation, some functions may be missing and are implemented anew. For example, in Hypre, distributed sparse matrices split into multiple stripes, where each MPI rank owns a stripe with continuous row indices. In gradient back-propagation, the transpose of the original matrix is usually needed and such functionalities are missing in Hyper.

Note that ADCME uses hybrid parallel computing models, i.e., a mixture of multithreading programs and MPI communication; therefore, when we talk about \textbf{one MPI processor}, it may contain multiple CPU cores. The MPI processors may also be distributed in different CPUs/nodes, which are interconnected via network communications. 

All the experiments are conducted on a cluster with 133 computational node, and each node has 32 cores in total (Intel(R) Xeon(R) CPU E5-2670 0 @ 2.60GHz, 32 GB RAM). The MPI programs are verified with serial programs.

\subsection{Transposition}\label{sect:transposition}

This example benchmarks transposition of a distributed sparse matrix. The transposition operation is very important for gradient back-propagation. We consider solving the Poisson's equation in $[0,1]^2$
\begin{equation}
	\begin{aligned}
		\nabla \cdot (\kappa(\bx) \nabla u(\bx)) & = f(\bx) & \bx \in \Omega \\ 
		u(\bx) &= 0 & \bx \in \partial \Omega
	\end{aligned}
\end{equation}
Here $f(\bx)\equiv 1$, and $\kappa(\bx)$ is approximated by a deep neural network 
$$\kappa(\bx)  = \mathcal{N}_\theta(\bx)$$
where $\theta$ is the neural network weights and biases. The equation is discretized using finite difference method on a uniform grid and the discretization leads to a linear system 
$$\mathbf{A}\mathbf{u} = \mathbf{f}$$
Here $\mathbf{u}$ is the solution vector and $\mathbf{f}$ is the source vector. Note that $\mathbf{A}$ is a sparse matrix and its entries depend on $\theta$. 

The domain is split into multiple patches and each patch correspond to a strip of rows in $\mathbf{A}$. The results are shown in \Cref{fig:transpose}. In the plots, we show the strong scaling for a fixed matrix size of 25,000,000$\times$25,000,000 as well as the weak scaling, where each MPI processor owns a patch of the whole mesh and has a size $300\times 300$, i.e., $300^2=90000$ rows. The run time for a fixed problem size is effectively reduced as we increase the number of processors, and no stagnation is observed for the current setting. For the weak scaling results, we see only 2$\sim$3 times increase in runtime when we scale up to hundreds of processors.

\begin{figure}[htpb]
	\centering 
	\scalebox{0.8}{
\begin{tikzpicture}

\definecolor{color0}{rgb}{0.12156862745098,0.466666666666667,0.705882352941177}
\definecolor{color1}{rgb}{1,0.498039215686275,0.0549019607843137}

\begin{axis}[
legend cell align={left},
legend style={fill opacity=0.8, draw opacity=1, text opacity=1, at={(0.03,0.03)}, anchor=south west, draw=white!80!black},
log basis x={10},
log basis y={10},
tick align=outside,
tick pos=left,
x grid style={white!69.0196078431373!black},
xlabel={Number of Processors},
xmajorgrids,
xmin=0.794328234724281, xmax=125.892541179417,
xmode=log,
xtick style={color=black},
xtick={0.01,0.1,1,10,100,1000,10000},
xticklabels={\(\displaystyle {10^{-2}}\),\(\displaystyle {10^{-1}}\),\(\displaystyle {10^{0}}\),\(\displaystyle {10^{1}}\),\(\displaystyle {10^{2}}\),\(\displaystyle {10^{3}}\),\(\displaystyle {10^{4}}\)},
y grid style={white!69.0196078431373!black},
ylabel={Time (sec)},
ymajorgrids,
ymin=0.000609773245147519, ymax=38.0612792528017,
ymode=log,
ytick style={color=black},
ytick={1e-05,0.0001,0.001,0.01,0.1,1,10,100,1000},
yticklabels={\(\displaystyle {10^{-5}}\),\(\displaystyle {10^{-4}}\),\(\displaystyle {10^{-3}}\),\(\displaystyle {10^{-2}}\),\(\displaystyle {10^{-1}}\),\(\displaystyle {10^{0}}\),\(\displaystyle {10^{1}}\),\(\displaystyle {10^{2}}\),\(\displaystyle {10^{3}}\)}
]
\addplot [semithick, color0, mark=*, mark size=3, mark options={solid}, forget plot]
table {%
1 23.0416999340057
4 8.50059998035431
16 2.1396999835968
25 1.39119999408722
64 0.592300009727478
100 0.402899956703186
};
\end{axis}

\end{tikzpicture}}~
	\scalebox{0.8}{
\begin{tikzpicture}

\definecolor{color0}{rgb}{0.12156862745098,0.466666666666667,0.705882352941177}

\begin{axis}[
log basis x={10},
tick align=outside,
tick pos=left,
x grid style={white!69.0196078431373!black},
xlabel={Number of Processors},
xmajorgrids,
xmin=0.794328234724281, xmax=125.892541179417,
xmode=log,
xtick style={color=black},
xtick={0.01,0.1,1,10,100,1000,10000},
xticklabels={\(\displaystyle {10^{-2}}\),\(\displaystyle {10^{-1}}\),\(\displaystyle {10^{0}}\),\(\displaystyle {10^{1}}\),\(\displaystyle {10^{2}}\),\(\displaystyle {10^{3}}\),\(\displaystyle {10^{4}}\)},
y grid style={white!69.0196078431373!black},
ylabel={Time (sec)},
ymajorgrids,
ymin=0.0, ymax=0.2,
ytick style={color=black},
ytick={0.05,0.075,0.1,0.125,0.15,0.175,0.2},
yticklabels={0.050,0.075,0.100,0.125,0.150,0.175,0.200}
]
\addplot [semithick, color0, mark=*, mark size=3, mark options={solid}]
table {%
1 0.0642999172210693
4 0.0936000108718872
16 0.0955999851226807
25 0.0954999923706055
36 0.116500043869019
49 0.0974000215530395
64 0.150399994850159
81 0.115399980545044
100 0.180099940299988
};
\end{axis}

\end{tikzpicture}}
	\caption{Strong (left) and weak (right) scaling for transposition of distributed sparse matrix.}
	\label{fig:transpose}
\end{figure}
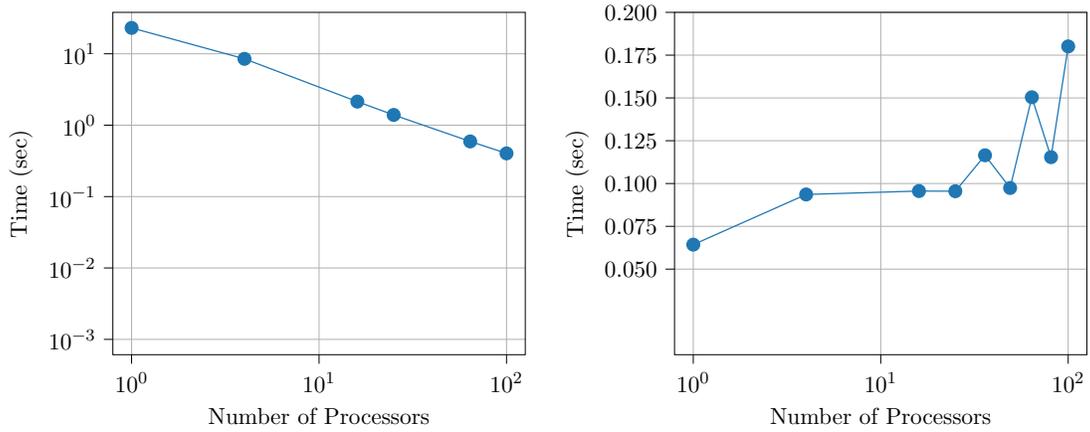

\subsection{Poisson's Equation}

\begin{figure}[htpb]
	\centering 
	\includegraphics[width=0.6\textwidth]{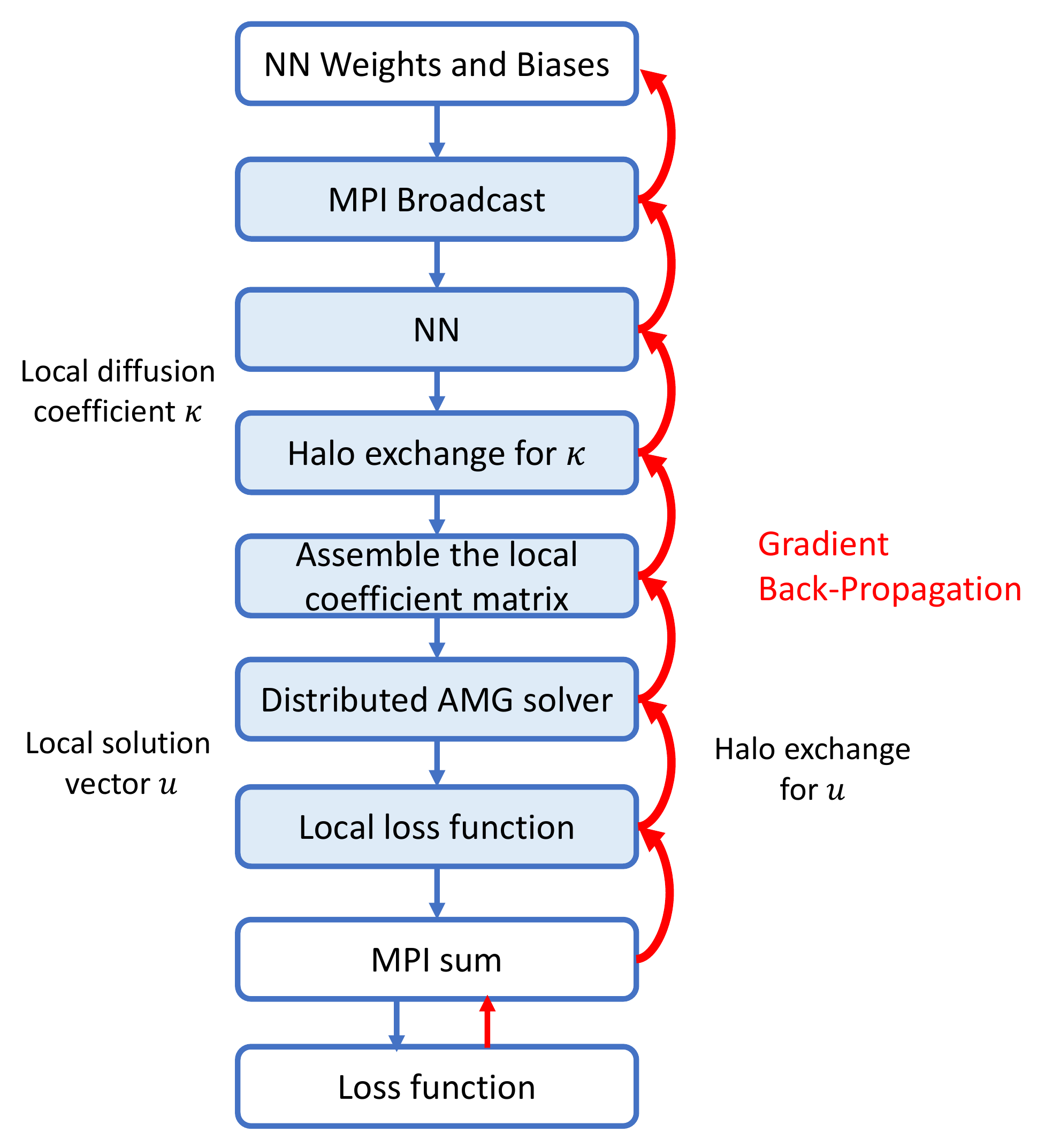}
	\caption{Parallel computing model for solving the Poisson's equation.}
	\label{fig:poisson}
\end{figure}

This example demonstrates the capability of ADCME to use Hypre for solving linear systems in parallel. We show that the overhead of ADCME is quite small compared to the linear solver time. We use the same problem setting as \Cref{sect:transposition}. We formulate the loss function using the numerical solution  
$$ J(\theta):=\sum_{i\in \mathcal{I}} \left( u(\bx_i) - u_i \right)^2$$
Here $\mathcal{I}$ consists of 80\% randomly picked degrees of freedom for each patch. \Cref{fig:poisson} shows the computational model for solving the Poisson's equation.

In the strong scaling experiments, we consider a fixed problem size $1800 \times 1800$ (mesh size, which implies the matrix size is around 32 million $\times$ 32 million). In the weak scaling experiments, each MPI processor owns a $300\times 300$ block. For example, a problem with 3600 processors has the problem size $90000\times 3600 \approx 0.3$ billion.

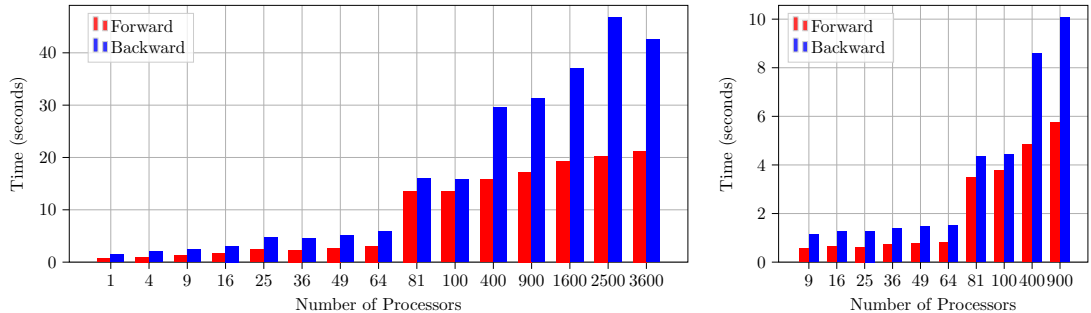
\begin{figure}[htpb]
	\centering 
	\scalebox{0.6}{
\begin{tikzpicture}

\begin{axis}[
legend cell align={left},
legend style={fill opacity=0.8, draw opacity=1, text opacity=1, at={(0.03,0.97)}, anchor=north west, draw=white!80!black},
tick align=outside,
tick pos=left,
x grid style={white!69.0196078431373!black},
xlabel={Number of Processors},
xmajorgrids,
xmin=-0.085, xmax=16.085,
xminorgrids,
xtick style={color=black},
xtick={1,2,3,4,5,6,7,8,9,10,11,12,13,14,15},
xticklabels={1,4,9,16,25,36,49,64,81,100,400,900,1600,2500,3600},
y grid style={white!69.0196078431373!black},
ylabel={Time (seconds)},
ymajorgrids,
ymin=0, ymax=49.067674142655,
yminorgrids,
ytick style={color=black},
x post scale=2.0
]
\draw[draw=none,fill=red] (axis cs:0.65,0) rectangle (axis cs:1,0.7502938682);
\addlegendimage{ybar,ybar legend,draw=none,fill=red};
\addlegendentry{Forward}

\draw[draw=none,fill=red] (axis cs:1.65,0) rectangle (axis cs:2,0.9904287783);
\draw[draw=none,fill=red] (axis cs:2.65,0) rectangle (axis cs:3,1.2721277562);
\draw[draw=none,fill=red] (axis cs:3.65,0) rectangle (axis cs:4,1.596422384);
\draw[draw=none,fill=red] (axis cs:4.65,0) rectangle (axis cs:5,2.3906420824);
\draw[draw=none,fill=red] (axis cs:5.65,0) rectangle (axis cs:6,2.3038286736);
\draw[draw=none,fill=red] (axis cs:6.65,0) rectangle (axis cs:7,2.7002926732);
\draw[draw=none,fill=red] (axis cs:7.65,0) rectangle (axis cs:8,3.0788769124);
\draw[draw=none,fill=red] (axis cs:8.65,0) rectangle (axis cs:9,13.5993086005);
\draw[draw=none,fill=red] (axis cs:9.65,0) rectangle (axis cs:10,13.5442737341);
\draw[draw=none,fill=red] (axis cs:10.65,0) rectangle (axis cs:11,15.8050156711);
\draw[draw=none,fill=red] (axis cs:11.65,0) rectangle (axis cs:12,17.2093361084);
\draw[draw=none,fill=red] (axis cs:12.65,0) rectangle (axis cs:13,19.3299712852);
\draw[draw=none,fill=red] (axis cs:13.65,0) rectangle (axis cs:14,20.2964213763);
\draw[draw=none,fill=red] (axis cs:14.65,0) rectangle (axis cs:15,21.2351118867);
\draw[draw=none,fill=blue] (axis cs:1,0) rectangle (axis cs:1.35,1.4139647905);
\addlegendimage{ybar,ybar legend,draw=none,fill=blue};
\addlegendentry{Backward}

\draw[draw=none,fill=blue] (axis cs:2,0) rectangle (axis cs:2.35,2.0312182473);
\draw[draw=none,fill=blue] (axis cs:3,0) rectangle (axis cs:3.35,2.4761697202);
\draw[draw=none,fill=blue] (axis cs:4,0) rectangle (axis cs:4.35,3.0733501376);
\draw[draw=none,fill=blue] (axis cs:5,0) rectangle (axis cs:5.35,4.7225032634);
\draw[draw=none,fill=blue] (axis cs:6,0) rectangle (axis cs:6.35,4.4907741909);
\draw[draw=none,fill=blue] (axis cs:7,0) rectangle (axis cs:7.35,5.1580138193);
\draw[draw=none,fill=blue] (axis cs:8,0) rectangle (axis cs:8.35,5.8528464565);
\draw[draw=none,fill=blue] (axis cs:9,0) rectangle (axis cs:9.35,15.9661503437);
\draw[draw=none,fill=blue] (axis cs:10,0) rectangle (axis cs:10.35,15.8007244653);
\draw[draw=none,fill=blue] (axis cs:11,0) rectangle (axis cs:11.35,29.6185663174);
\draw[draw=none,fill=blue] (axis cs:12,0) rectangle (axis cs:12.35,31.3219851481);
\draw[draw=none,fill=blue] (axis cs:13,0) rectangle (axis cs:13.35,37.062870086);
\draw[draw=none,fill=blue] (axis cs:14,0) rectangle (axis cs:14.35,46.7311182311);
\draw[draw=none,fill=blue] (axis cs:15,0) rectangle (axis cs:15.35,42.5510425847);
\end{axis}

\end{tikzpicture}}~
	\scalebox{0.6}{
\begin{tikzpicture}

\begin{axis}[
legend cell align={left},
legend style={fill opacity=0.8, draw opacity=1, text opacity=1, at={(0.03,0.97)}, anchor=north west, draw=white!80!black},
tick align=outside,
tick pos=left,
x grid style={white!69.0196078431373!black},
xlabel={Number of  Processors},
xmin=0.165, xmax=11,
xmajorgrids,
xmin=-0.085, xmax=11,
xminorgrids,
xtick style={color=black},
xtick={1,2,3,4,5,6,7,8,9,10},
xticklabels={9,16,25,36,49,64,81,100,400,900},
y grid style={white!69.0196078431373!black},
ylabel={Time (seconds)},
ymajorgrids,
ymin=0, ymax=12,
yminorgrids,
ymin=0, ymax=10.567210069185,
ytick style={color=black}
]
\draw[draw=none,fill=red] (axis cs:0.65,0) rectangle (axis cs:1,0.5535179348);
\addlegendimage{ybar,ybar legend,draw=none,fill=red};
\addlegendentry{Forward}

\draw[draw=none,fill=red] (axis cs:1.65,0) rectangle (axis cs:2,0.6566348317);
\draw[draw=none,fill=red] (axis cs:2.65,0) rectangle (axis cs:3,0.621591817);
\draw[draw=none,fill=red] (axis cs:3.65,0) rectangle (axis cs:4,0.7202479615);
\draw[draw=none,fill=red] (axis cs:4.65,0) rectangle (axis cs:5,0.7632927056);
\draw[draw=none,fill=red] (axis cs:5.65,0) rectangle (axis cs:6,0.8001853406);
\draw[draw=none,fill=red] (axis cs:6.65,0) rectangle (axis cs:7,3.4809433753);
\draw[draw=none,fill=red] (axis cs:7.65,0) rectangle (axis cs:8,3.7753181108);
\draw[draw=none,fill=red] (axis cs:8.65,0) rectangle (axis cs:9,4.8336867255);
\draw[draw=none,fill=red] (axis cs:9.65,0) rectangle (axis cs:10,5.755047674);
\draw[draw=none,fill=blue] (axis cs:1,0) rectangle (axis cs:1.35,1.1463683659);
\addlegendimage{ybar,ybar legend,draw=none,fill=blue};
\addlegendentry{Backward}

\draw[draw=none,fill=blue] (axis cs:2,0) rectangle (axis cs:2.35,1.2659345722);
\draw[draw=none,fill=blue] (axis cs:3,0) rectangle (axis cs:3.35,1.2849252896);
\draw[draw=none,fill=blue] (axis cs:4,0) rectangle (axis cs:4.35,1.3713638365);
\draw[draw=none,fill=blue] (axis cs:5,0) rectangle (axis cs:5.35,1.4558442149);
\draw[draw=none,fill=blue] (axis cs:6,0) rectangle (axis cs:6.35,1.5125004645);
\draw[draw=none,fill=blue] (axis cs:7,0) rectangle (axis cs:7.35,4.3344616447);
\draw[draw=none,fill=blue] (axis cs:8,0) rectangle (axis cs:8.35,4.4342483421);
\draw[draw=none,fill=blue] (axis cs:9,0) rectangle (axis cs:9.35,8.5910793756);
\draw[draw=none,fill=blue] (axis cs:10,0) rectangle (axis cs:10.35,10.0640095897);
\end{axis}

\end{tikzpicture}}
	\caption{Runtime for forward computation and gradient back-propagation. Left: each MPI processor has 1 core; right: each MPI processor has 4 cores. We see a jump near 64 cores because this is where network communications start taking place (recall each of our CPUs has 32 cores, and each node has two CPUs). The plots show results of weak scaling, each MPI processor solves a local problem of the same size}
	\label{fig:weak_poisson}
\end{figure}

We first consider the weak scaling. \Cref{fig:weak_poisson} shows the runtime for forward computation as well as gradient back-propagation. There are two important observations:
\begin{enumerate}
	\item By using more cores per processor, the runtime is reduced significantly. For example, the runtime for the backward is reduced to around 10 seconds from 30 seconds by switching from 1 core to 4 cores per processor.
	\item The runtime for the backward is typically less than twice the forward computation. Although the backward requires solve two linear systems (one of them is in the forward computation), the AMG (algebraic multigrid) linear solver in the back-propagation may converge faster, and therefore costs less than the forward.
\end{enumerate}

\begin{figure}[htpb]
	\centering 
	\scalebox{0.6}{
\begin{tikzpicture}

\begin{axis}[
legend cell align={left},
legend style={fill opacity=0.8, draw opacity=1, text opacity=1, at={(0.03,0.97)}, anchor=north west, draw=white!80!black},
tick align=outside,
tick pos=left,
x grid style={white!69.0196078431373!black},
xlabel={Number of Processors},
xmin=-0.085, xmax=16.085,
xtick style={color=black},
xtick={1,2,3,4,5,6,7,8,9,10,11,12,13,14,15},
xticklabels={1,4,9,16,25,36,49,64,81,100,400,900,1600,2500,3600},
y grid style={white!69.0196078431373!black},
ylabel={Time (seconds)},
ymin=0, ymax=0.757224470909705,
ytick style={color=black},
ytick={0,0.1,0.2,0.3,0.4,0.5,0.6,0.7,0.8},
yticklabels={0.0,0.1,0.2,0.3,0.4,0.5,0.6,0.7,0.8},
yminorgrids,
ymajorgrids,
xminorgrids,
xmajorgrids,
x post scale=2.0
]
\draw[draw=none,fill=red] (axis cs:0.65,0) rectangle (axis cs:1,0.0230518990914185);
\addlegendimage{ybar,ybar legend,draw=none,fill=red};
\addlegendentry{Forward}

\draw[draw=none,fill=red] (axis cs:1.65,0) rectangle (axis cs:2,0.0481887785288817);
\draw[draw=none,fill=red] (axis cs:2.65,0) rectangle (axis cs:3,0.100991758427783);
\draw[draw=none,fill=red] (axis cs:3.65,0) rectangle (axis cs:4,0.146490690922912);
\draw[draw=none,fill=red] (axis cs:4.65,0) rectangle (axis cs:5,0.224442255205786);
\draw[draw=none,fill=red] (axis cs:5.65,0) rectangle (axis cs:6,0.17313688301925);
\draw[draw=none,fill=red] (axis cs:6.65,0) rectangle (axis cs:7,0.224601688474048);
\draw[draw=none,fill=red] (axis cs:7.65,0) rectangle (axis cs:8,0.259193241402381);
\draw[draw=none,fill=red] (axis cs:8.65,0) rectangle (axis cs:9,0.232537951543701);
\draw[draw=none,fill=red] (axis cs:9.65,0) rectangle (axis cs:10,0.255760264403953);
\draw[draw=none,fill=red] (axis cs:10.65,0) rectangle (axis cs:11,0.293099987353662);
\draw[draw=none,fill=red] (axis cs:11.65,0) rectangle (axis cs:12,0.314553827478061);
\draw[draw=none,fill=red] (axis cs:12.65,0) rectangle (axis cs:13,0.324299879407764);
\draw[draw=none,fill=red] (axis cs:13.65,0) rectangle (axis cs:14,0.35278770072474);
\draw[draw=none,fill=red] (axis cs:14.65,0) rectangle (axis cs:15,0.310511690815297);
\draw[draw=none,fill=blue] (axis cs:1,0) rectangle (axis cs:1.35,0.0436848823197633);
\addlegendimage{ybar,ybar legend,draw=none,fill=blue};
\addlegendentry{Backward}

\draw[draw=none,fill=blue] (axis cs:2,0) rectangle (axis cs:2.35,0.0939984129723022);
\draw[draw=none,fill=blue] (axis cs:3,0) rectangle (axis cs:3.35,0.181086316612658);
\draw[draw=none,fill=blue] (axis cs:4,0) rectangle (axis cs:4.35,0.260045784839685);
\draw[draw=none,fill=blue] (axis cs:5,0) rectangle (axis cs:5.35,0.455584484980505);
\draw[draw=none,fill=blue] (axis cs:6,0) rectangle (axis cs:6.35,0.332923639294775);
\draw[draw=none,fill=blue] (axis cs:7,0) rectangle (axis cs:7.35,0.444031809412305);
\draw[draw=none,fill=blue] (axis cs:8,0) rectangle (axis cs:8.35,0.586310578318542);
\draw[draw=none,fill=blue] (axis cs:9,0) rectangle (axis cs:9.35,0.486402499957629);
\draw[draw=none,fill=blue] (axis cs:10,0) rectangle (axis cs:10.35,0.447165424276744);
\draw[draw=none,fill=blue] (axis cs:11,0) rectangle (axis cs:11.35,0.591402883371376);
\draw[draw=none,fill=blue] (axis cs:12,0) rectangle (axis cs:12.35,0.606468270925621);
\draw[draw=none,fill=blue] (axis cs:13,0) rectangle (axis cs:13.35,0.656457722947628);
\draw[draw=none,fill=blue] (axis cs:14,0) rectangle (axis cs:14.35,0.721166162771148);
\draw[draw=none,fill=blue] (axis cs:15,0) rectangle (axis cs:15.35,0.67584000043334);
\end{axis}

\end{tikzpicture}}~
	\scalebox{0.6}{
\begin{tikzpicture}

\begin{axis}[
legend cell align={left},
legend style={fill opacity=0.8, draw opacity=1, text opacity=1, at={(0.5,0.91)}, anchor=north, draw=white!80!black},
tick align=outside,
tick pos=left,
x grid style={white!69.0196078431373!black},
xlabel={Number of Processors},
xmin=0.165, xmax=10.835,
xtick style={color=black},
xtick={1,2,3,4,5,6,7,8,9,10},
xticklabels={9,16,25,36,49,64,81,100,400,900},
y grid style={white!69.0196078431373!black},
ylabel={Time (seconds)},
ymin=0, ymax=0.173020593410235,
ytick style={color=black},
yminorgrids,
ymajorgrids,
xminorgrids,
xmajorgrids
]
\draw[draw=none,fill=red] (axis cs:0.65,0) rectangle (axis cs:1,0.0577573747642944);
\addlegendimage{ybar,ybar legend,draw=none,fill=red};
\addlegendentry{Forward}

\draw[draw=none,fill=red] (axis cs:1.65,0) rectangle (axis cs:2,0.0826086523863403);
\draw[draw=none,fill=red] (axis cs:2.65,0) rectangle (axis cs:3,0.0776603089357299);
\draw[draw=none,fill=red] (axis cs:3.65,0) rectangle (axis cs:4,0.0810623421457519);
\draw[draw=none,fill=red] (axis cs:4.65,0) rectangle (axis cs:5,0.0810663099930052);
\draw[draw=none,fill=red] (axis cs:5.65,0) rectangle (axis cs:6,0.0879668414163451);
\draw[draw=none,fill=red] (axis cs:6.65,0) rectangle (axis cs:7,0.0882227951989867);
\draw[draw=none,fill=red] (axis cs:7.65,0) rectangle (axis cs:8,0.0830670245649543);
\draw[draw=none,fill=red] (axis cs:8.65,0) rectangle (axis cs:9,0.0854451340464468);
\draw[draw=none,fill=red] (axis cs:9.65,0) rectangle (axis cs:10,0.0871010015604252);
\draw[draw=none,fill=blue] (axis cs:1,0) rectangle (axis cs:1.35,0.111988788263281);
\addlegendimage{ybar,ybar legend,draw=none,fill=blue};
\addlegendentry{Backward}

\draw[draw=none,fill=blue] (axis cs:2,0) rectangle (axis cs:2.35,0.15562921522063);
\draw[draw=none,fill=blue] (axis cs:3,0) rectangle (axis cs:3.35,0.13695690199624);
\draw[draw=none,fill=blue] (axis cs:4,0) rectangle (axis cs:4.35,0.15593255159552);
\draw[draw=none,fill=blue] (axis cs:5,0) rectangle (axis cs:5.35,0.141555455668432);
\draw[draw=none,fill=blue] (axis cs:6,0) rectangle (axis cs:6.35,0.149073588431885);
\draw[draw=none,fill=blue] (axis cs:7,0) rectangle (axis cs:7.35,0.156801942398975);
\draw[draw=none,fill=blue] (axis cs:8,0) rectangle (axis cs:8.35,0.152526225629664);
\draw[draw=none,fill=blue] (axis cs:9,0) rectangle (axis cs:9.35,0.159577200268885);
\draw[draw=none,fill=blue] (axis cs:10,0) rectangle (axis cs:10.35,0.164781517533557);
\end{axis}

\end{tikzpicture}}
	\caption{Runtime for forward computation and gradient back-propagation. The plots show results of weak scaling, each MPI processor solves a local problem of the same size. However, each MPI processor may have 1 or 4 cores.}
	\label{fig:poisson_overhead}
\end{figure}
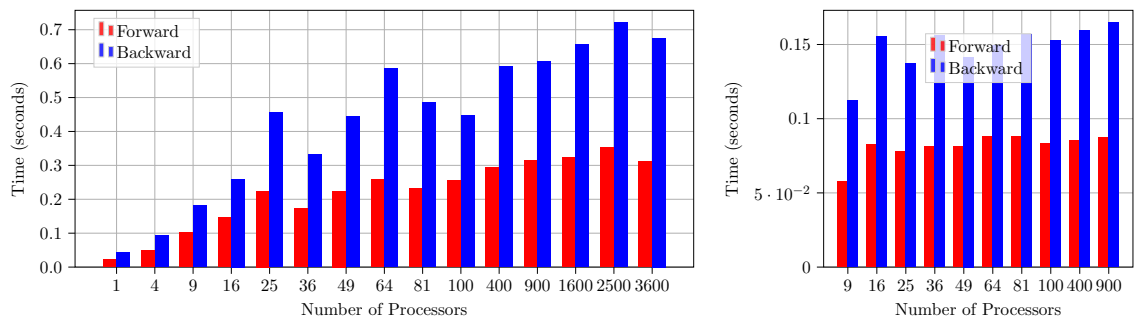

Additionally, we show the overhead in \Cref{fig:poisson_overhead}, which is defined as the difference between total runtime and Hypre linear solver time, of both the forward and backward calculation.

We see that the overhead is quite small compared to the total time, especially when the problem size is large. This indicates that the ADCME MPI implementation is very effective.

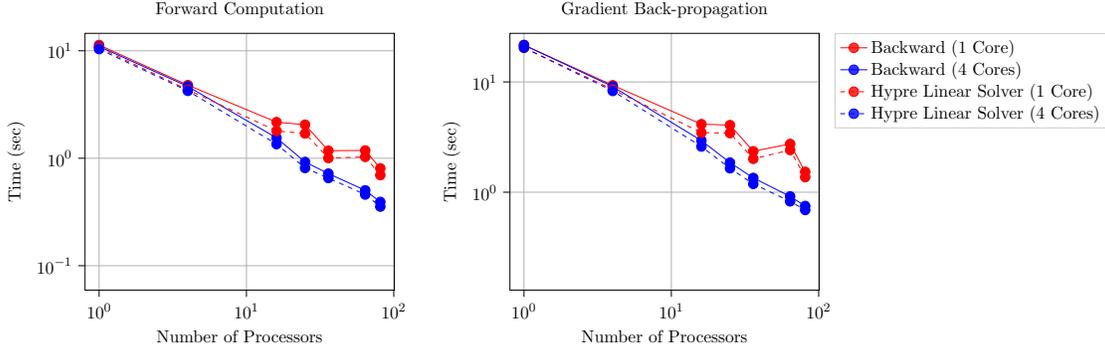
\begin{figure}[htpb]
	\centering 
	\scalebox{0.6}{
\begin{tikzpicture}

\begin{axis}[
legend cell align={left},
legend style={fill opacity=0.8, draw opacity=1, text opacity=1, at={(1.05,1)}, anchor=north west, draw=white!80!black},
log basis x={10},
log basis y={10},
tick align=outside,
tick pos=left,
title={Forward Computation},
x grid style={white!69.0196078431373!black},
xlabel={Number of Processors},
xmajorgrids,
xmin=0.802741561760231, xmax=100.904206108857,
xminorgrids,
xmode=log,
xtick style={color=black},
xtick={0.01,0.1,1,10,100,1000,10000},
xticklabels={\(\displaystyle {10^{-2}}\),\(\displaystyle {10^{-1}}\),\(\displaystyle {10^{0}}\),\(\displaystyle {10^{1}}\),\(\displaystyle {10^{2}}\),\(\displaystyle {10^{3}}\),\(\displaystyle {10^{4}}\)},
y grid style={white!69.0196078431373!black},
ylabel={Time (sec)},
ymajorgrids,
ymin=0.0593014710967097, ymax=14.5376257665934,
yminorgrids,
ymode=log,
ytick style={color=black},
ytick={0.001,0.01,0.1,1,10,100,1000},
yticklabels={\(\displaystyle {10^{-3}}\),\(\displaystyle {10^{-2}}\),\(\displaystyle {10^{-1}}\),\(\displaystyle {10^{0}}\),\(\displaystyle {10^{1}}\),\(\displaystyle {10^{2}}\),\(\displaystyle {10^{3}}\)}
]
\addplot [semithick, red, mark=*, mark size=3, mark options={solid}]
table {%
1 11.3209564683333
4 4.79422741966667
16 2.17050120933333
25 2.051652212
36 1.17613905466667
64 1.18216338966667
81 0.805914490666667
};

\addplot [semithick, blue, mark=*, mark size=3, mark options={solid}]
table {%
1 10.9740334753333
4 4.65844348766667
16 1.54950077633333
25 0.923378885
36 0.719816114666667
64 0.503020210333333
81 0.392305410333333
};

\addplot [semithick, red, dashed, mark=*, mark size=3, mark options={solid}]
table {%
1 10.7133537928263
4 4.3389786084493
16 1.79605770111084
25 1.70004932085673
36 1.00369930267334
64 1.02990531921387
81 0.696585257848104
};

\addplot [semithick, blue, dashed, mark=*, mark size=3, mark options={solid}]
table {%
1 10.3704266548157
4 4.24105739593506
16 1.3537962436676
25 0.81137760480245
36 0.654465595881144
64 0.459558327992757
81 0.355130116144816
};

\end{axis}

\end{tikzpicture}}~
	\scalebox{0.6}{
\begin{tikzpicture}

\begin{axis}[
legend cell align={left},
legend style={fill opacity=0.8, draw opacity=1, text opacity=1, at={(1.05,1)}, anchor=north west, draw=white!80!black},
log basis x={10},
log basis y={10},
tick align=outside,
tick pos=left,
title={Gradient Back-propagation},
x grid style={white!69.0196078431373!black},
xlabel={Number of Processors},
xmajorgrids,
xmin=0.802741561760231, xmax=100.904206108857,
xminorgrids,
xmode=log,
xtick style={color=black},
xtick={0.01,0.1,1,10,100,1000,10000},
xticklabels={\(\displaystyle {10^{-2}}\),\(\displaystyle {10^{-1}}\),\(\displaystyle {10^{0}}\),\(\displaystyle {10^{1}}\),\(\displaystyle {10^{2}}\),\(\displaystyle {10^{3}}\),\(\displaystyle {10^{4}}\)},
y grid style={white!69.0196078431373!black},
ylabel={Time (sec)},
ymajorgrids,
ymin=0.129506945891268, ymax=27.5782682687816,
yminorgrids,
ymode=log,
ytick style={color=black},
ytick={0.01,0.1,1,10,100,1000},
yticklabels={\(\displaystyle {10^{-2}}\),\(\displaystyle {10^{-1}}\),\(\displaystyle {10^{0}}\),\(\displaystyle {10^{1}}\),\(\displaystyle {10^{2}}\),\(\displaystyle {10^{3}}\)}
]
\addplot [semithick, red, mark=*, mark size=3, mark options={solid}]
table {%
1 21.5627918103333
4 9.326442247
16 4.14030789866667
25 4.059420941
36 2.35263105666667
64 2.738277528
81 1.52939861333333
};
\addlegendentry{Backward (1 Core)}
\addplot [semithick, blue, mark=*, mark size=3, mark options={solid}]
table {%
1 21.614060298
4 9.07585462833333
16 2.93765755333333
25 1.85637265666667
36 1.34677541866667
64 0.916292855666667
81 0.750087024666667
};
\addlegendentry{Backward (4 Cores)}
\addplot [semithick, red, dashed, mark=*, mark size=3, mark options={solid}]
table {%
1 20.4045383135478
4 8.53607535362244
16 3.46877233187358
25 3.4349213441213
36 2.00878969828288
64 2.40959326426188
81 1.36863533655802
};
\addlegendentry{Hypre Linear Solver
(1 Core)}
\addplot [semithick, blue, dashed, mark=*, mark size=3, mark options={solid}]
table {%
1 20.4419123331706
4 8.28150312105815
16 2.59433237711589
25 1.64779241879781
36 1.1886150042216
64 0.826372623443604
81 0.689682722091675
};
\addlegendentry{Hypre Linear Solver
(4 Cores)}

\end{axis}

\end{tikzpicture}}
	\caption{Runtime for forward computation and gradient back-propagation. The plots show results of strong scaling, the whole problem size is fixed and we increase the number of MPI processors (each processor has 1 or 4 cores).}
	\label{fig:strong_poisson}
\end{figure}

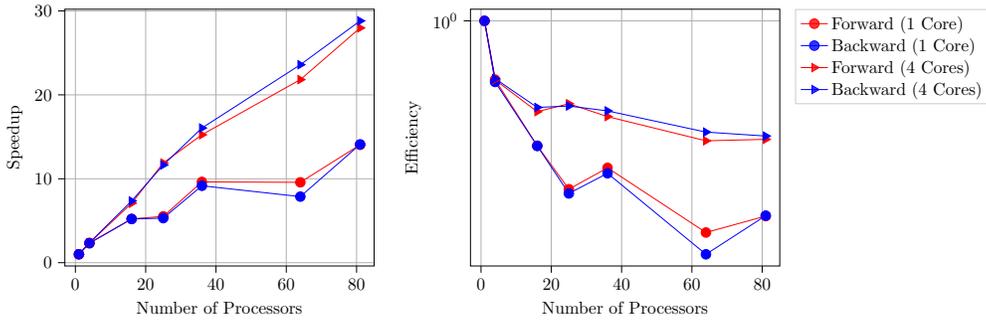
\begin{figure}[htpb]
	\centering 
	\scalebox{0.6}{
\begin{tikzpicture}

\begin{axis}[
legend cell align={left},
legend style={fill opacity=0.8, draw opacity=1, text opacity=1, at={(1.05,1)}, anchor=north west, draw=white!80!black},
tick align=outside,
tick pos=left,
x grid style={white!69.0196078431373!black},
xlabel={Number of Processors},
xmajorgrids,
xmin=-3, xmax=85,
xminorgrids,
xtick style={color=black},
y grid style={white!69.0196078431373!black},
ylabel={Speedup},
ymajorgrids,
ymin=-0.390770176474199, ymax=30.2061737059582,
yminorgrids,
ytick style={color=black}
]
\addplot [semithick, red, mark=*, mark size=3, mark options={solid}]
table {%
1 1
4 2.36137243341712
16 5.21582592059949
25 5.51797054204299
36 9.62552550518089
64 9.57647357995538
81 14.0473419940228
};

\addplot [semithick, blue, mark=*, mark size=3, mark options={solid}]
table {%
1 1
4 2.31200614760354
16 5.20801649009663
25 5.31179006161937
36 9.16539452679191
64 7.87458232039851
81 14.0988697271911
};

\addplot [semithick, red, mark=triangle*, mark size=3, mark options={solid,rotate=270}]
table {%
1 1
4 2.35572965613672
16 7.08230266350803
25 11.8846484943538
36 15.2456068317048
64 21.8162873974015
81 27.9731892201256
};

\addplot [semithick, blue, mark=triangle*, mark size=3, mark options={solid,rotate=270}]
table {%
1 1
4 2.38149035910342
16 7.3575833484998
25 11.6431688542593
36 16.0487487359981
64 23.5885941534208
81 28.815403529484
};

\end{axis}

\end{tikzpicture}}~
		\scalebox{0.6}{
\begin{tikzpicture}

\begin{axis}[
legend cell align={left},
legend style={fill opacity=0.8, draw opacity=1, text opacity=1, at={(1.05,1)}, anchor=north west, draw=white!80!black},
log basis y={10},
tick align=outside,
tick pos=left,
x grid style={white!69.0196078431373!black},
xlabel={Number of Processors},
xmajorgrids,
xmin=-3, xmax=85,
xminorgrids,
xtick style={color=black},
y grid style={white!69.0196078431373!black},
ylabel={Efficiency},
ymajorgrids,
ymin=0.110802594855376, ymax=1.11044645585082,
yminorgrids,
ymode=log,
ytick style={color=black},
ytick={0.01,0.1,1,10,100},
yticklabels={\(\displaystyle {10^{-2}}\),\(\displaystyle {10^{-1}}\),\(\displaystyle {10^{0}}\),\(\displaystyle {10^{1}}\),\(\displaystyle {10^{2}}\)}
]
\addplot [semithick, red, mark=*, mark size=3, mark options={solid}]
table {%
1 1
4 0.590343108354279
16 0.325989120037468
25 0.220718821681719
36 0.267375708477247
64 0.149632399686803
81 0.173423975234849
};
\addlegendentry{Forward (1 Core)}
\addplot [semithick, blue, mark=*, mark size=3, mark options={solid}]
table {%
1 1
4 0.578001536900884
16 0.32550103063104
25 0.212471602464775
36 0.254594292410886
64 0.123040348756227
81 0.174060120088779
};
\addlegendentry{Backward (1 Core)}
\addplot [semithick, red, mark=triangle*, mark size=3, mark options={solid,rotate=270}]
table {%
1 1
4 0.58893241403418
16 0.442643916469252
25 0.475385939774152
36 0.423489078658466
64 0.340879490584398
81 0.345348015063278
};
\addlegendentry{Forward (4 Cores)}
\addplot [semithick, blue, mark=triangle*, mark size=3, mark options={solid,rotate=270}]
table {%
1 1
4 0.595372589775855
16 0.459848959281238
25 0.465726754170374
36 0.445798575999948
64 0.3685717836472
81 0.355745722586222
};
\addlegendentry{Backward (4 Cores)}
\end{axis}

\end{tikzpicture}}
	\caption{Speedup and efficiency for parallel computing of the Poisson's equation.}
	\label{fig:speedup_efficiency_poisson}
\end{figure}

Now we consider the strong scaling. In this case, we fixed the whole problem size and split the mesh onto different MPI processors. \Cref{fig:strong_poisson} shows the runtime for the forward computation and the gradient back-propagation. We can reduce the runtime by more than 20 times for the expensive gradient back-propagation by utilizing more than 100 MPI processors. \Cref{fig:speedup_efficiency_poisson} shows the speedup and efficiency. We can see that the 4 cores have smaller runtime compared to 1 core.

\begin{figure}[htpb]
	\centering 
	\includegraphics[width=0.8\textwidth]{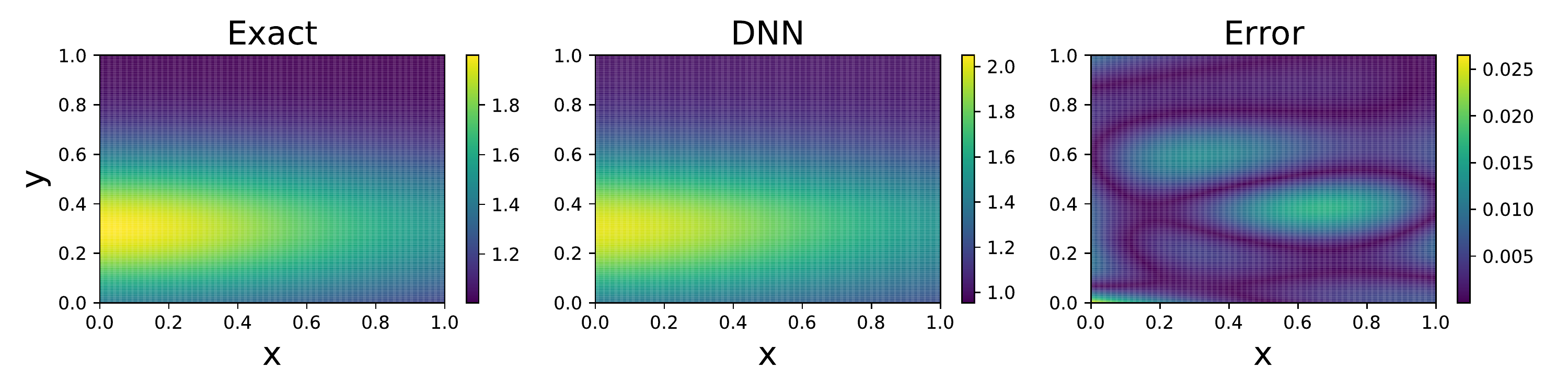}
	\includegraphics[width=0.8\textwidth]{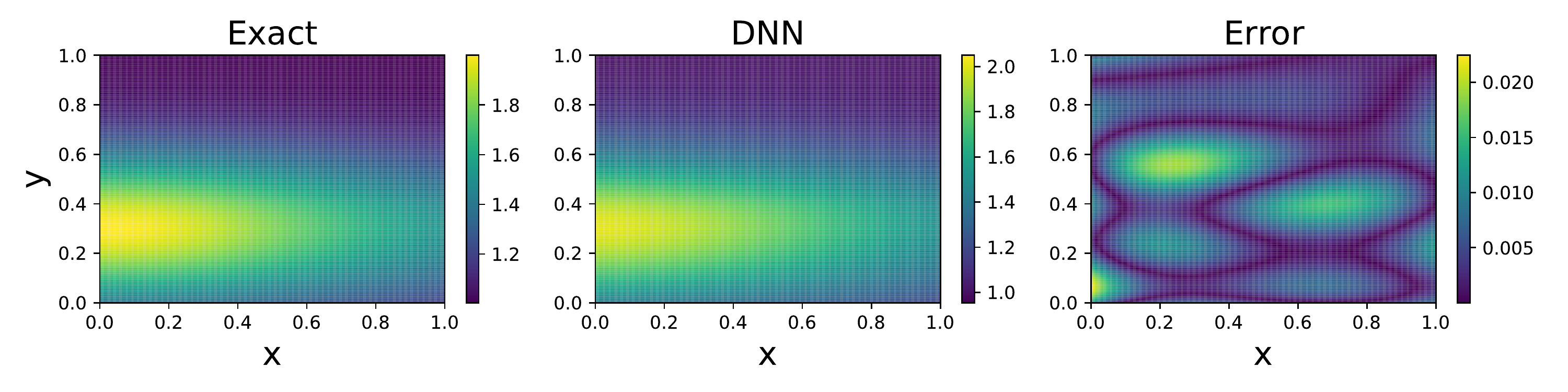}
	\caption{Estimated $\kappa(\bx)$ at 400-th iteration. The first row shows the result with 4 MPI processors (the mesh size is $600\times 600$), and the second row shows the result with 9 MPI processors (the mesh size is $900\times 900$).}
	\label{fig:N4N9_pdf}
\end{figure}

\begin{figure}[htpb]
	\centering 
	\scalebox{0.6}{
\begin{tikzpicture}

\definecolor{color0}{rgb}{0.12156862745098,0.466666666666667,0.705882352941177}

\begin{axis}[
log basis y={10},
tick align=outside,
tick pos=left,
x grid style={white!69.0196078431373!black},
xlabel={Iteration},
xmajorgrids,
xmin=-20, xmax=420,
xtick style={color=black},
y grid style={white!69.0196078431373!black},
ylabel={Loss},
ymajorgrids,
ymin=397119.738060744, ymax=148275817816.883,
ymode=log,
ytick style={color=black},
ytick={10000,100000,1000000,10000000,100000000,1000000000,10000000000,100000000000,1000000000000,10000000000000},
yticklabels={\(\displaystyle {10^{4}}\),\(\displaystyle {10^{5}}\),\(\displaystyle {10^{6}}\),\(\displaystyle {10^{7}}\),\(\displaystyle {10^{8}}\),\(\displaystyle {10^{9}}\),\(\displaystyle {10^{10}}\),\(\displaystyle {10^{11}}\),\(\displaystyle {10^{12}}\),\(\displaystyle {10^{13}}\)}
]
\addplot [semithick, color0]
table {%
0 82754350541.6372
1 78213845903.1838
2 76222791019.5282
3 72682371209.4097
4 44201564075.517
5 30872472488.2415
6 22794930948.7371
7 10163770034.8109
8 7624169413.51
9 6565782325.96953
10 4088858617.24046
11 3787651041.07051
12 3624609593.6435
13 3611521255.56724
14 3279527876.30244
15 3115734140.30323
16 2969305316.86922
17 2915425535.05378
18 2796286547.41432
19 2654529359.36528
20 2532624958.3102
21 2437886614.94252
22 2311186872.16537
23 2234374031.47156
24 2082612170.07352
25 2041375386.85663
26 1990085010.69535
27 1974951968.01184
28 1971698109.70105
29 1969592121.31656
30 1958312202.39811
31 1702211331.2394
32 1549614206.57719
33 1423025197.60091
34 1341412274.23609
35 1274893316.79106
36 1137072319.28224
37 1048114893.36304
38 924940698.723084
39 829345452.764806
40 650946491.398354
41 530761179.206219
42 499958304.109222
43 497697468.040089
44 493737286.981893
45 489080911.057782
46 479593339.401371
47 452665594.037856
48 409862398.196585
49 384529523.211275
50 361287369.329994
51 340045887.129723
52 291545314.195718
53 285594747.470366
54 256381643.142517
55 245251410.498549
56 224633365.175945
57 205461820.325818
58 166389719.230046
59 138702430.4658
60 124339410.043458
61 101540481.646362
62 98546609.1467403
63 95491657.7118842
64 92613235.7359751
65 84504939.3100923
66 78033921.7418048
67 76738617.685976
68 74843429.8703104
69 62300986.8636873
70 51334647.8963125
71 42557960.733544
72 41373291.7510059
73 41137760.5269839
74 40966652.94575
75 40850780.2261873
76 40292887.8124471
77 39635701.8980426
78 37239509.4341201
79 30079270.0924554
80 22802005.5493819
81 18475416.3441175
82 15082621.6698858
83 12244369.4416816
84 10954256.9808189
85 10475158.399507
86 9795468.90315788
87 9157862.92606114
88 8752296.69022562
89 8320783.97300132
90 7478481.87209695
91 7418998.77971232
92 7004333.47100277
93 6880739.80295815
94 6826704.08076459
95 6796560.70136623
96 6766477.1547986
97 6758349.56824983
98 6754912.77818528
99 6750254.29083709
100 6744485.47023338
101 6731058.05422332
102 6625611.702106
103 6489959.8308658
104 6391314.50681208
105 6361939.41186414
106 6333499.65996581
107 6311707.42923818
108 6294971.34380188
109 6285799.15058218
110 6277979.55852535
111 6266582.89607107
112 6255297.69228512
113 6223731.07024641
114 6197507.18254121
115 6170057.27778822
116 6164551.88700908
117 6163461.37193042
118 6160131.58067161
119 6155319.72471811
120 6153002.69115155
121 6149959.85342844
122 6147967.35113349
123 6142392.01062457
124 6136406.00959274
125 6135132.69339529
126 6134054.26375223
127 6128536.00722844
128 6108335.36443888
129 6053776.56661503
130 5986899.73866664
131 5920951.12499646
132 5778382.62379314
133 5697804.86285909
134 5576930.46872168
135 5563639.56171346
136 5430697.09119425
137 5366693.59177032
138 5310147.61504453
139 5192862.14743431
140 5021597.01323744
141 4761691.1663241
142 4440633.4789211
143 4327383.90228728
144 4279971.44422836
145 4261296.85936022
146 4240499.11704137
147 4233059.26400338
148 4230268.61299266
149 4212302.78039526
150 4200951.90749432
151 4195197.09799028
152 4192180.91237962
153 4181447.5699991
154 4164966.32588915
155 4158236.20622393
156 4155663.84954259
157 4150944.18875025
158 4141633.76446879
159 4135884.72714902
160 4133928.76860141
161 4130867.80627813
162 4119932.83529583
163 4117976.13247252
164 4114201.46287526
165 4111724.8215368
166 4108772.86846116
167 4107624.02633212
168 4107497.35888794
169 4107257.19195572
170 4107149.41079819
171 4103748.4665208
172 4101771.71274313
173 4099271.49934378
174 3930515.87210443
175 3725398.2303913
176 3509145.08061231
177 3358581.4372576
178 3338191.00998331
179 3312487.23649537
180 3281971.46289168
181 3237915.15902795
182 3212541.04786114
183 3190664.08589818
184 3178212.106489
185 3154293.1810404
186 3144753.46062012
187 3134954.82308594
188 3132753.92730328
189 3128830.8052006
190 3126240.62893005
191 3125643.73735836
192 3125045.04243717
193 3123601.72198441
194 3122886.38997974
195 3122702.2792114
196 3122634.5030589
197 3122351.84676216
198 3121105.80725077
199 3115570.73773837
200 3115145.53590429
201 3115018.0848938
202 3114112.07655558
203 3113326.56693245
204 3112267.4468201
205 3110578.43867092
206 3091840.32239829
207 3073942.56619605
208 3056878.93577356
209 3041148.07553643
210 3026902.71009119
211 2928061.6604598
212 2906767.23748993
213 2822510.60619999
214 2730159.12862277
215 2610175.46477487
216 2445306.53805599
217 2390984.91343655
218 2347397.75766164
219 2335508.60505117
220 2288371.03111367
221 2203478.22539948
222 2162910.23429517
223 2150830.94595526
224 2140068.86373818
225 2125269.34576589
226 2109001.11828348
227 2065080.48917964
228 2043287.13740272
229 2012928.31511767
230 1943837.12795733
231 1896390.55054949
232 1851402.31067471
233 1804875.79872698
234 1794194.40179038
235 1786741.04659167
236 1783881.15618198
237 1778584.21979218
238 1769427.99598929
239 1755017.78966773
240 1703974.79544009
241 1684863.57054907
242 1680157.9393311
243 1677825.81956522
244 1670774.31190601
245 1668235.60331549
246 1667212.87807638
247 1663829.63913882
248 1659690.68525776
249 1632715.53038916
250 1599125.55509778
251 1578193.44010212
252 1572310.05193286
253 1568143.25829343
254 1562892.15468216
255 1560989.57913998
256 1549460.17600599
257 1547727.13187266
258 1547431.1293974
259 1547179.39212843
260 1545976.43686643
261 1544919.56477776
262 1544344.89014022
263 1543977.77229795
264 1542356.7252032
265 1541928.14839751
266 1541751.89951213
267 1541411.00700658
268 1539904.63049427
269 1534741.53278021
270 1531375.49230576
271 1530495.68081034
272 1529373.27575907
273 1528119.77797871
274 1524389.70644531
275 1523552.63080501
276 1521242.48107417
277 1519333.48294587
278 1517862.43278598
279 1516721.38460053
280 1515072.07577215
281 1503942.15934795
282 1497546.64455105
283 1491318.41805115
284 1482277.42799255
285 1474659.31782517
286 1469279.93923081
287 1463446.78655838
288 1461075.94328804
289 1460039.85716766
290 1458068.66932052
291 1452599.19339546
292 1449800.35319541
293 1445088.78949444
294 1442124.79667906
295 1434230.04793963
296 1429159.0735402
297 1427637.87033733
298 1426573.08219889
299 1418726.76248371
300 1375909.97963951
301 1323987.8798494
302 1316608.76567258
303 1313582.94137212
304 1305846.076686
305 1299154.17032094
306 1296264.72393014
307 1291775.15972573
308 1285321.88717914
309 1277699.81061781
310 1274182.34186668
311 1267901.32155431
312 1260703.52903032
313 1256210.68574388
314 1253368.3468614
315 1251714.32491751
316 1250628.56002742
317 1250423.22448029
318 1250214.29592122
319 1249590.95426698
320 1249312.29126002
321 1248688.16796053
322 1246741.38528079
323 1233220.27020787
324 1224932.60824545
325 1218085.82861935
326 1216329.19460716
327 1214215.79220693
328 1210868.8384325
329 1208524.07398288
330 1206949.37319309
331 1204809.4649654
332 1199356.83613035
333 1184232.37433271
334 1169825.99309686
335 1162674.72994985
336 1154407.13280277
337 1153690.33724446
338 1151081.23632627
339 1149226.25634661
340 1147489.03993405
341 1142632.49528102
342 1126936.42101272
343 1119977.00663558
344 1113136.84049487
345 1098198.8869451
346 1093494.89314983
347 1071016.54136451
348 1064768.8264557
349 1059796.46919227
350 1058783.91451516
351 1056706.68668809
352 1055922.80840376
353 1045608.34356203
354 1039488.13209403
355 1033476.72407275
356 1027335.7180654
357 1022987.70628344
358 1014240.58861365
359 1006196.10058804
360 1001775.34589493
361 995342.442266591
362 988338.136241139
363 977975.260803666
364 958356.13297324
365 949790.413507312
366 944925.615944063
367 942510.21455834
368 937865.177900839
369 928023.388742251
370 920938.300948471
371 915541.552637212
372 905747.910034855
373 892821.178324557
374 860210.773280462
375 844172.835043314
376 827407.516037436
377 810378.645645747
378 769464.850262178
379 759953.20318264
380 755684.337194627
381 750018.493464057
382 749323.888290342
383 746504.200381463
384 745421.876558909
385 743100.041234003
386 740829.765291426
387 736778.491165006
388 734354.011248
389 733386.249031627
390 730893.882783981
391 730117.173322911
392 729248.863766884
393 727295.130351634
394 725862.283634371
395 725747.112952077
396 725690.347218126
397 725364.829762838
398 724055.122373796
399 718050.714826979
400 711542.698925013
};
\end{axis}

\end{tikzpicture}}~
	\scalebox{0.6}{
\begin{tikzpicture}

\definecolor{color0}{rgb}{0.12156862745098,0.466666666666667,0.705882352941177}

\begin{axis}[
log basis y={10},
tick align=outside,
tick pos=left,
x grid style={white!69.0196078431373!black},
xlabel={Iteration},
xmajorgrids,
xmin=-20, xmax=420,
xtick style={color=black},
y grid style={white!69.0196078431373!black},
ylabel={Loss},
ymajorgrids,
ymin=2018022.37375163, ymax=320553527895.019,
ymode=log,
ytick style={color=black},
ytick={100000,1000000,10000000,100000000,1000000000,10000000000,100000000000,1000000000000,10000000000000},
yticklabels={\(\displaystyle {10^{5}}\),\(\displaystyle {10^{6}}\),\(\displaystyle {10^{7}}\),\(\displaystyle {10^{8}}\),\(\displaystyle {10^{9}}\),\(\displaystyle {10^{10}}\),\(\displaystyle {10^{11}}\),\(\displaystyle {10^{12}}\),\(\displaystyle {10^{13}}\)}
]
\addplot [semithick, color0]
table {%
0 185991297999.01
1 176031078143.07
2 171420736468.095
3 164907823133.39
4 90157696914.3811
5 68324094888.4641
6 48653671706.334
7 20299636319.1947
8 17448615801.6717
9 16760500079.8898
10 15597824494.207
11 11824279694.6731
12 8670372353.18858
13 8393455596.35744
14 8388247520.74998
15 7922360936.31837
16 7320421190.42359
17 6993213190.96545
18 6617496128.65349
19 6231665934.26869
20 5818156148.31614
21 5603024249.17536
22 5350185135.8984
23 5061172205.55524
24 4765927782.74115
25 4569446386.68147
26 4482631514.43979
27 4434555072.50037
28 4403222513.28311
29 4399348741.99045
30 4367430241.85948
31 3723158276.14471
32 3421034280.84047
33 3292577201.04462
34 2908691467.17928
35 2493388595.6226
36 2438308110.17218
37 2252875904.97489
38 1997734909.28758
39 1753610191.4919
40 1620835050.68306
41 1497252442.06576
42 1460988448.19609
43 1455197292.40442
44 1427832143.20482
45 1241865674.34785
46 1055982694.33932
47 903197900.403807
48 799882122.730421
49 717627565.772974
50 672794241.139745
51 624567280.590914
52 611525016.197545
53 591543732.580557
54 571572535.974034
55 559339461.266886
56 552085607.566767
57 508595956.167103
58 456073278.23342
59 445440548.94192
60 427459214.719176
61 400201241.391213
62 360297671.853117
63 331869525.085042
64 292453562.838135
65 277441558.116673
66 258618751.722305
67 226802689.292732
68 191876065.457254
69 162440595.346436
70 152811694.44959
71 128217731.355493
72 77325092.8553851
73 73531969.7970319
74 71799794.3503586
75 70162918.7917675
76 68413058.2604526
77 67764745.9592552
78 66649867.4888033
79 64729218.5214866
80 64233900.3576454
81 63993960.3989143
82 63180289.5326356
83 62051028.0732423
84 58725093.9262391
85 55261810.3080998
86 51401520.8147413
87 49533445.4915895
88 45328303.2274181
89 44241815.0639407
90 43561025.9751039
91 43244186.8272458
92 42772704.2635111
93 41524543.6282639
94 41381180.3794502
95 40726904.7873291
96 40436808.6257313
97 40037309.1458594
98 39843835.6969466
99 39699618.3061895
100 39610948.7038874
101 39357391.0999655
102 39196281.5676111
103 39104559.5818809
104 38761119.5679083
105 38361114.3429351
106 37920223.3189252
107 37464319.5248217
108 37269607.3837844
109 37191006.0226328
110 37151816.2360444
111 37097606.3279644
112 37070526.7098747
113 37043767.6472842
114 36975831.0061058
115 36896435.4799239
116 36855518.6138564
117 36575893.9876961
118 36061560.6049006
119 28998063.2048968
120 27285216.0131616
121 25335304.77047
122 23054999.7032922
123 22763391.1504971
124 21362129.9813958
125 21176154.1788983
126 20114055.3510201
127 20000698.2139849
128 19295853.7434354
129 19035401.0220566
130 18968536.1562094
131 18919834.71855
132 18813661.3604145
133 18773202.5946577
134 18378503.2331966
135 18333786.0089102
136 18246696.4917262
137 18150817.9752123
138 18110735.4245773
139 18103005.5853372
140 18092366.7385323
141 18084376.6736557
142 18075467.3813559
143 18049762.6685594
144 18014299.2042262
145 18003353.8396393
146 18002860.1149616
147 17999620.3399016
148 17934999.7735581
149 17861675.4436386
150 17855681.2061135
151 17820821.0457353
152 17784557.6873502
153 17759381.8578703
154 17557201.1846602
155 17464345.8210553
156 16786553.4772651
157 16298314.1936803
158 15748308.4692083
159 15371397.1561998
160 14996704.9424776
161 14241366.8041786
162 13733683.0897147
163 13508444.9882164
164 13400760.4601108
165 13262780.25324
166 13150500.1082524
167 12997483.2714711
168 12907047.7596045
169 12836649.7033017
170 12778386.3503572
171 12641634.8468857
172 12586147.0721793
173 12276726.6103928
174 12227133.2902249
175 12089471.6111522
176 12008271.6477841
177 11971574.3270066
178 11958962.0906667
179 11952452.0054978
180 11942861.2332079
181 11932168.3548377
182 11920208.3741329
183 11909164.439467
184 11904875.8094457
185 11899932.5135591
186 11895693.4914465
187 11892398.9917643
188 11891247.0040215
189 11889358.1626948
190 11886595.1984855
191 11883767.1424399
192 11882168.4900309
193 11879717.200812
194 11878282.5576316
195 11876891.6681119
196 11876137.773471
197 11874599.1097854
198 11870150.9501066
199 11859331.2430119
200 11854400.120794
201 11849807.7272494
202 11842960.7716215
203 11839176.1293287
204 11814021.6505625
205 11812389.8820002
206 11779961.7955448
207 11755213.6638276
208 11530373.8694821
209 11392553.1434037
210 11235503.208418
211 11114734.4748209
212 11042889.4368339
213 10978093.4609734
214 10949316.4126313
215 10914562.5139372
216 10889748.0435952
217 10859621.2176187
218 10831766.7326454
219 10817209.4442423
220 10805064.2572245
221 10784481.3228383
222 10748552.9839938
223 10743240.3064105
224 10578479.3816718
225 10551467.6968681
226 10487110.1015444
227 10453393.6243963
228 10345108.8406654
229 10274244.0707516
230 10248898.6830538
231 10199428.2420246
232 10185095.2583592
233 10169771.4781319
234 10135915.5722508
235 10071111.4064733
236 10024574.6491396
237 9953135.3157899
238 9879312.00868761
239 9748239.63625117
240 9643536.52079894
241 9559236.30728452
242 9494571.96391759
243 9461021.60500013
244 9426611.17002263
245 9416492.04455869
246 9394350.10605567
247 9373117.92933301
248 9343138.4141523
249 9218801.93843021
250 9158276.62488928
251 9111208.21349723
252 9053989.73877385
253 9039981.04874027
254 9029861.34131589
255 8970729.20968832
256 8937112.06814483
257 8924669.24297014
258 8921501.11227152
259 8920325.71443195
260 8915602.68961184
261 8913091.10149053
262 8910310.54619248
263 8909525.27774338
264 8908991.79780133
265 8908622.62984541
266 8897515.11858684
267 8887882.47604995
268 8866421.29418865
269 8854288.24818697
270 8849128.00706209
271 8844719.23110669
272 8842318.81786146
273 8833924.94948173
274 8801489.28528173
275 8703376.28056988
276 8590671.28696812
277 8551767.71140521
278 8528623.52439597
279 8493763.55774826
280 8390191.77082803
281 8321464.48432568
282 8278772.99941812
283 8252775.38991046
284 8210406.72578691
285 8194992.15590634
286 8160160.86066517
287 8141075.04586097
288 8101362.72854055
289 8073444.64362676
290 8049209.27035656
291 8014144.46003382
292 7972507.03710563
293 7836813.54614818
294 7636607.89620638
295 7493463.76932905
296 7301354.91425573
297 7178650.28060503
298 7064174.70081191
299 6996459.48196884
300 6944420.17697068
301 6863633.64508429
302 6741691.61601613
303 6566809.36873136
304 6520640.11740458
305 6393456.72288894
306 6260760.40589188
307 6184082.15378359
308 6059266.96415578
309 5959496.79958457
310 5841247.46420792
311 5745761.82096463
312 5712270.95957276
313 5646952.07528144
314 5531754.555285
315 5457113.58258814
316 5370856.06169812
317 5352203.50732204
318 5331893.52995678
319 5282260.99415331
320 5219322.25713576
321 5164357.26129668
322 5117241.72933006
323 5066677.85291391
324 5036062.82227212
325 5003567.89703315
326 4940512.37535553
327 4861588.23802236
328 4829697.94899344
329 4815348.26660487
330 4776266.3585731
331 4677755.88461542
332 4532865.81912212
333 4481004.77689923
334 4376945.88733516
335 4320672.28682423
336 4296108.51812686
337 4267755.94963956
338 4236796.09403592
339 4211922.60768663
340 4200633.63426168
341 4186174.54281626
342 4171913.30187816
343 4152596.87156092
344 4140861.35733867
345 4118652.44915105
346 4108321.97847287
347 4091292.90430697
348 4081675.77838874
349 4078358.47767735
350 4066867.02326292
351 4059939.63851369
352 4049332.14506027
353 4043961.79584614
354 4039681.6708584
355 4033031.89426429
356 4025774.13719636
357 4019195.50213289
358 3999606.52418208
359 3995066.27506673
360 3979469.16643309
361 3978346.31571613
362 3965611.91739496
363 3960612.02766075
364 3947089.25921836
365 3935989.44475132
366 3928181.17643247
367 3920230.05675156
368 3919141.97321976
369 3918592.01591037
370 3915586.67819832
371 3913847.86181618
372 3909208.54277307
373 3907780.7369178
374 3898331.33958521
375 3887220.78343024
376 3876825.6720928
377 3860820.16824045
378 3856836.94343022
379 3842626.58794559
380 3831775.46438047
381 3829658.43482369
382 3824187.08490843
383 3812157.62964236
384 3802006.90681166
385 3796043.64473837
386 3782203.18276678
387 3739747.26077161
388 3734946.21249573
389 3718885.16405024
390 3707163.81311355
391 3698329.86707243
392 3687212.22044131
393 3664260.32391561
394 3632313.98087767
395 3614377.430709
396 3590254.06064538
397 3546453.15895589
398 3498404.94072697
399 3485424.29527041
400 3478034.71579951
};
\end{axis}

\end{tikzpicture}}
	\caption{Loss function profiles for distributed optimization with 4 MPI processors (left) and 9 MPI processors (right).}
	\label{fig:N4N9}
\end{figure}
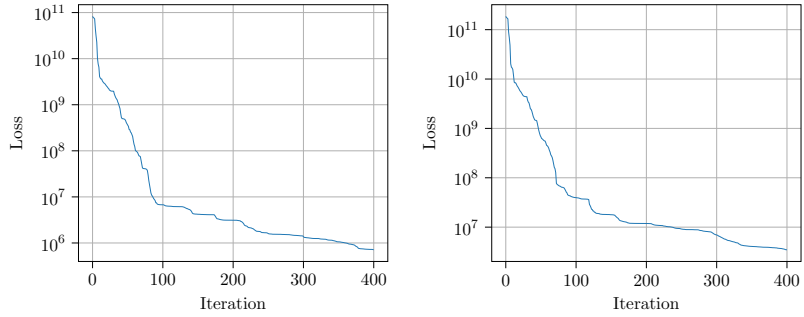

Finally, we apply the distributed L-BFGS optimizer to train the deep neural network. The distributed optimizer is constructed from a serial distributed optimizer using the approach described in \Cref{sect:distopt}. \Cref{fig:N4N9_pdf} illustrate the results at 400-th iteration with 4 and 9 MPI processors respectively. Each processor owns a $300\times 300$ patch. We can see that the estimated DNN-based $\kappa(\bx)$ is very similar to the exact $\kappa(\bx)$. \Cref{fig:N4N9} also shows the corresponding loss function profiles.

\subsection{Acoustic Wave Equation}

\begin{figure}[htpb]
	\centering 
	\includegraphics[width=0.18\textwidth]{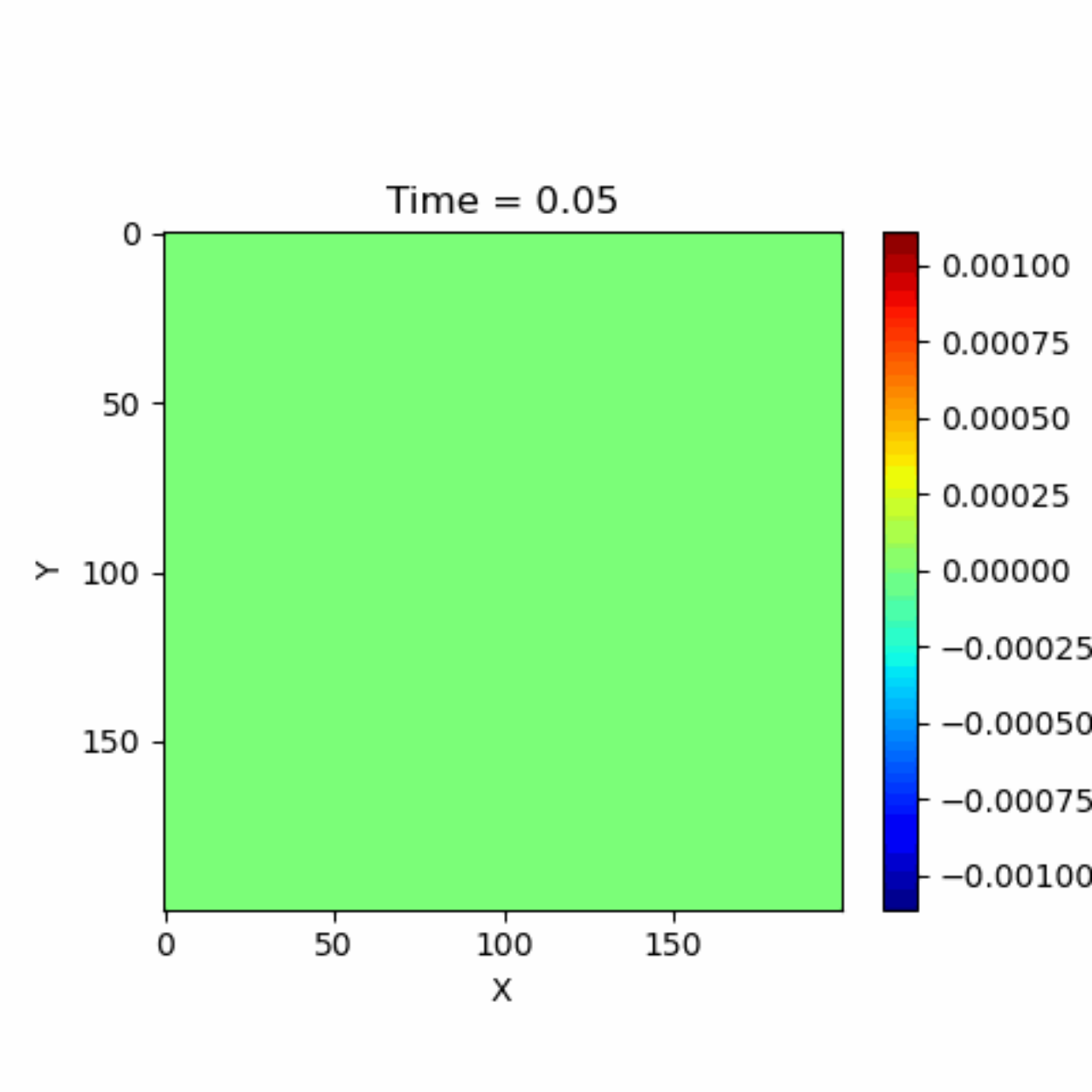}~
	\includegraphics[width=0.18\textwidth]{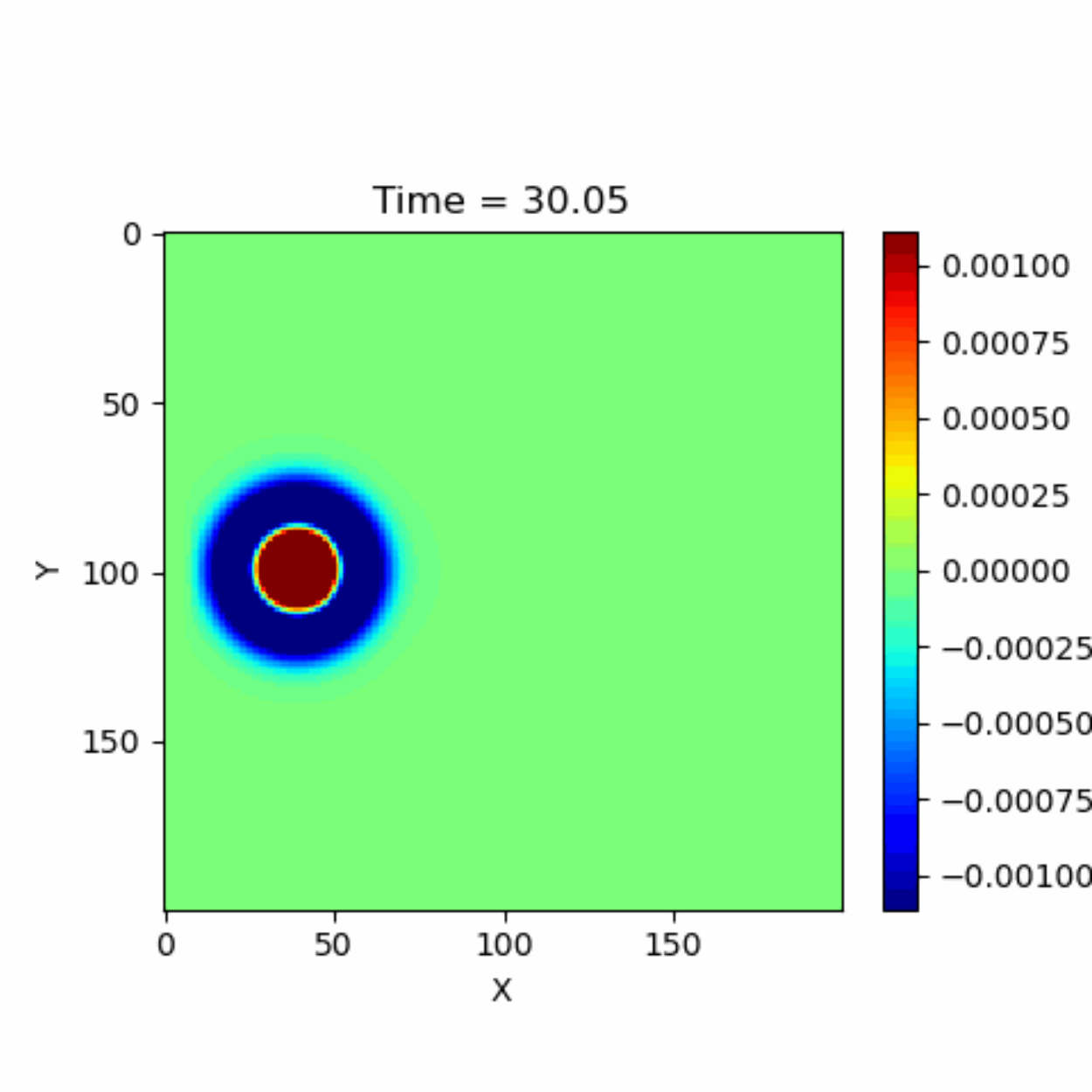}~
	\includegraphics[width=0.18\textwidth]{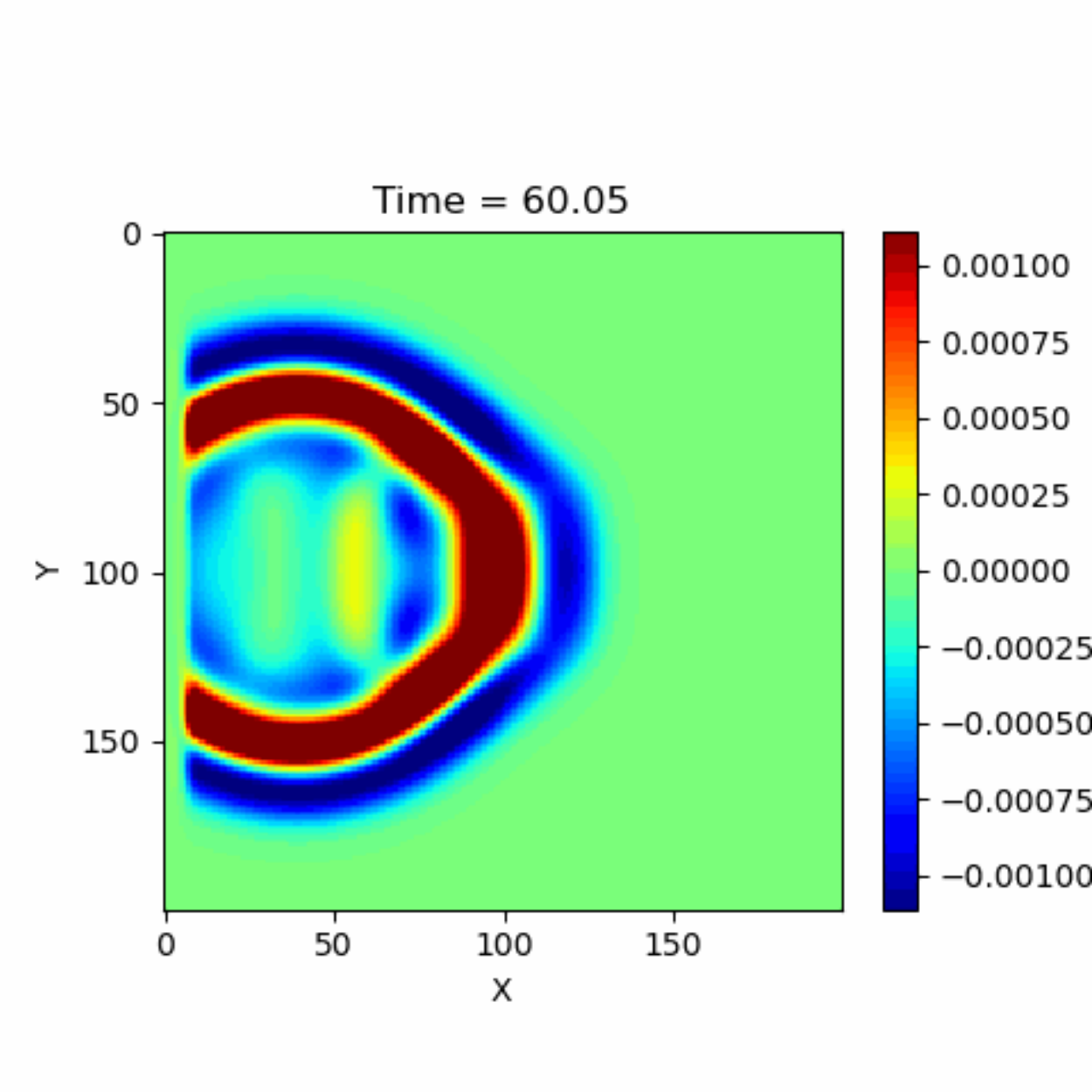}~
	\includegraphics[width=0.18\textwidth]{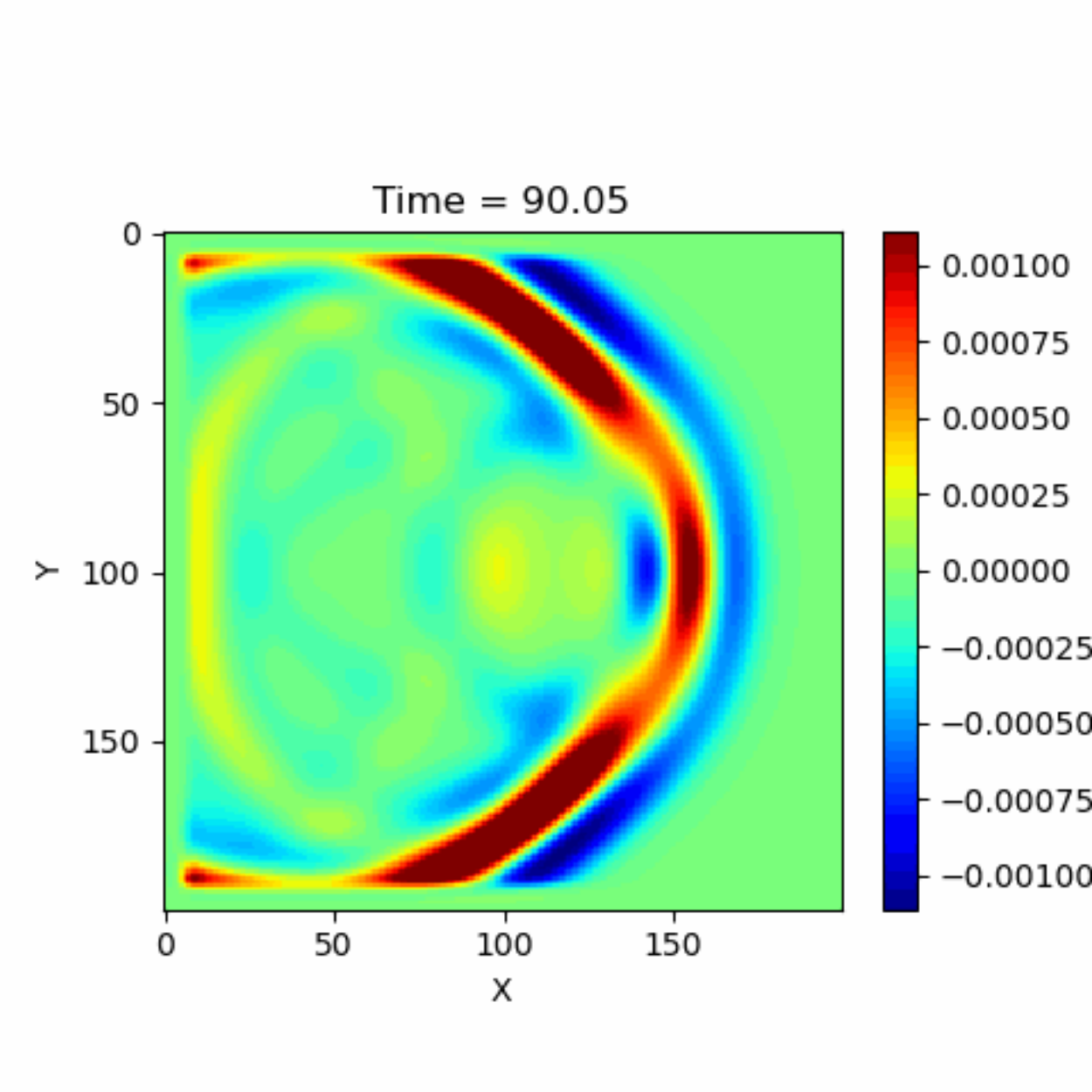}~
	\includegraphics[width=0.18\textwidth]{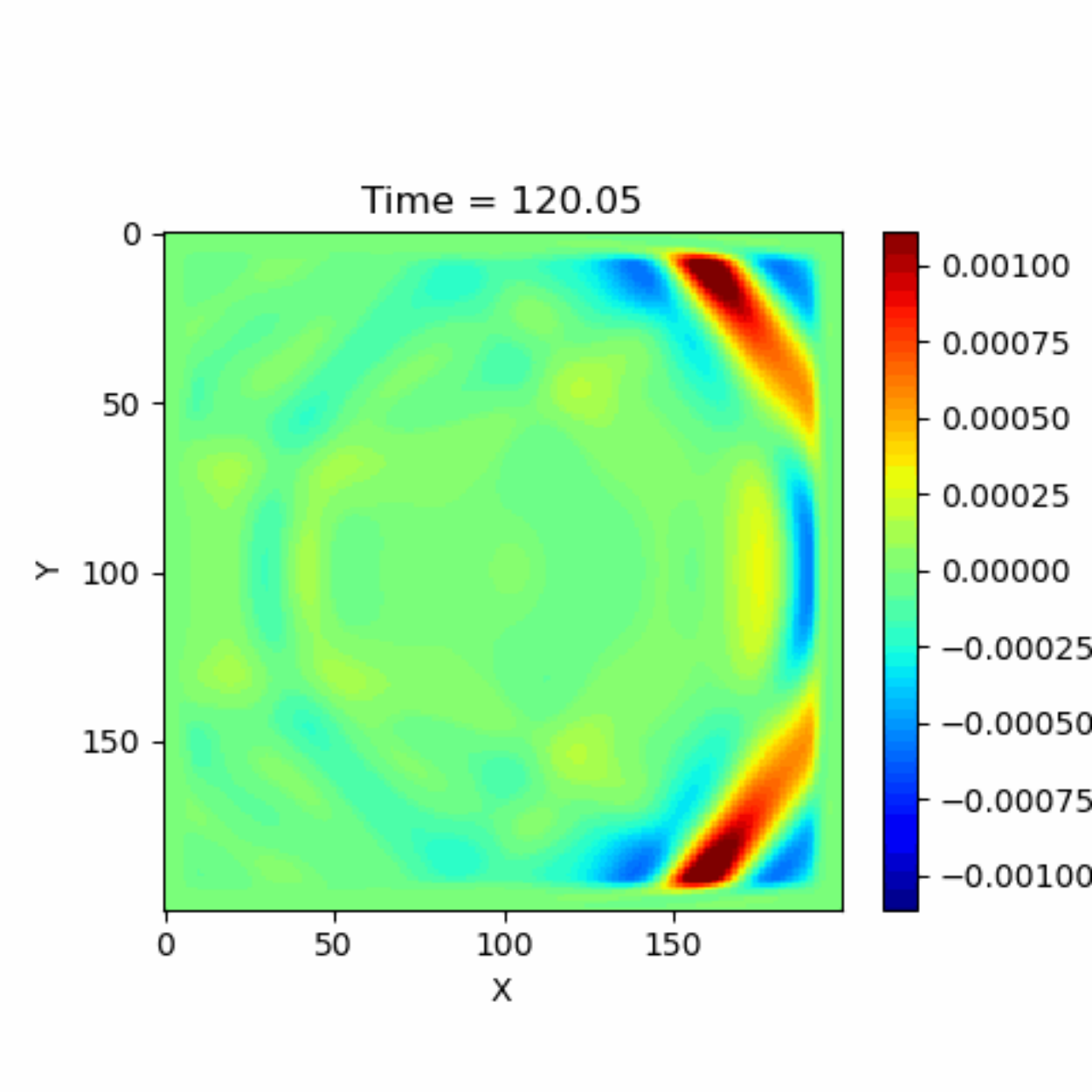}
	\caption{Snapshots of $v_1$ for the acoustic wave propagation.}
	\label{fig:acoustic_snapshots}
\end{figure}

\begin{figure}[htpb]
	\centering 
	\includegraphics[width=0.4\textwidth]{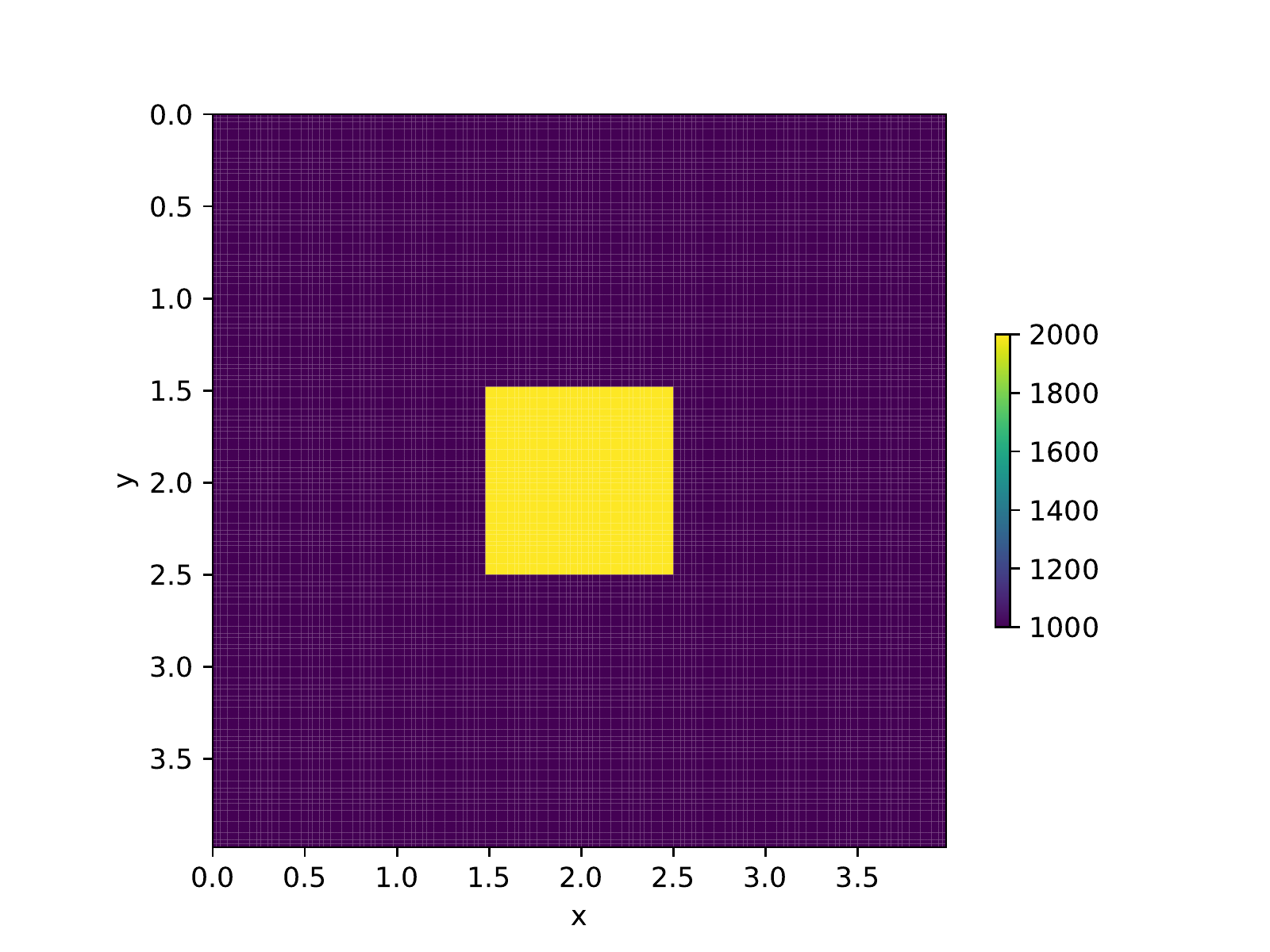}
	\caption{True velocity model $c(\bx)$.}
	\label{fig:acoustic_model}
\end{figure}

In this example, we consider the acoustic wave equation with perfectly matched layer (PML) \cite{grote2010efficient,zhu2020general}. The governing equation for the acoustic equation is
$$\frac{\partial^2 u}{\partial t^2} = \nabla \cdot (c^2 \nabla u)$$
where $u$ is the displacement, $f$ is the source term, and $c$ is the spatially varying acoustic velocity. \Cref{fig:acoustic_snapshots} shows the snapshots for the acoustic wave propagation and \Cref{fig:acoustic_model} shows the true velocity model we have used in the acoustic and elastic wave equation examples.

In the inverse problem, only the wavefield $u$ on the surface is observable, and we want to use this information to estimate $c$. The problem is usually ill-posed, so regularization techniques are usually used to constrain $c$. One approach is to represent $c$ by a deep neural network
$$c(x,y) = \mathcal{N}_\theta(x,y)$$
where $\theta$ is the neural network weights and biases. The loss function is formulated by the square loss for the wavefield on the surface. The computational model is shown in \Cref{fig:acoustic}.

\begin{figure}[htpb]
	\centering 
	\includegraphics[width=0.8\textwidth]{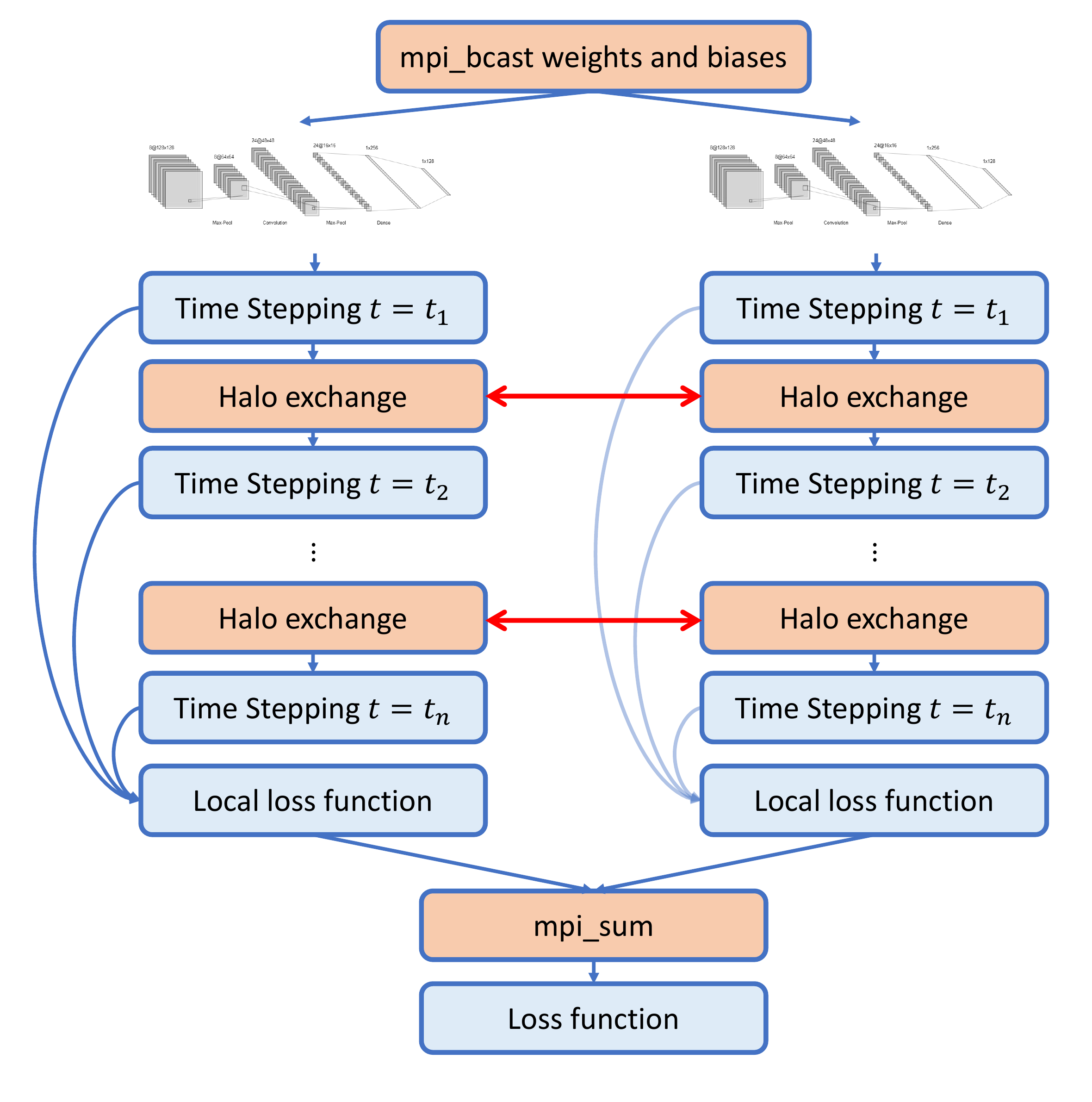}
	\caption{Parallel computing model for solving the acoustic wave equation.}
	\label{fig:acoustic}
\end{figure}

To implement an MPI version of the acoustic wave equation propagator, we use \texttt{mpi\_halo\_exchange}, which is implemented using MPI and performs the halo exchange mentioned in the last example for both wavefields and axilliary fields. This function communicates the boundary information for each block of the mesh. The following plot shows the computational graph for the numerical simulation of the acoustic wave equation

\Cref{fig:acoustic} shows the strong scaling and weak scaling of our implementation. Each processor consists of 32 processors, which are used at the discretion of ADCME's backend, i.e., TensorFlow. The strong scaling result is obtained by using a $1000\times 1000$ grid and 100 times steps. For the weak scaling result, each MPI processor owns a $100\times 100$ grid, and the total number of steps is 2000. It is remarkable that even though we increase the number of processors from 1 to 100, the total time only increases 2 times in the weak scaling.

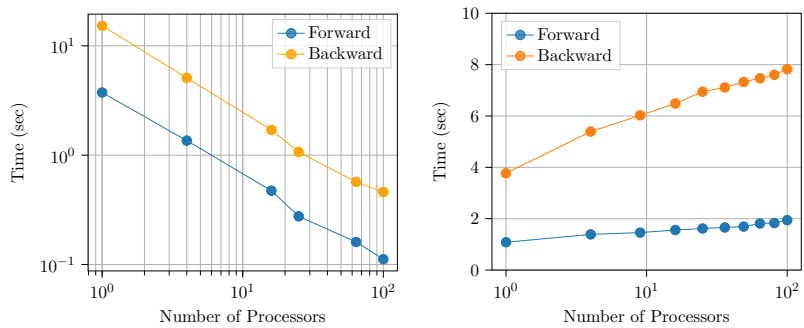
\begin{figure}[htpb]
	\centering 
	\scalebox{0.6}{
\begin{tikzpicture}

\definecolor{color0}{rgb}{0.12156862745098,0.466666666666667,0.705882352941177}
\definecolor{color1}{rgb}{1,0.647058823529412,0}

\begin{axis}[
legend cell align={left},
legend style={fill opacity=0.8, draw opacity=1, text opacity=1, draw=white!80!black},
log basis x={10},
log basis y={10},
tick align=outside,
tick pos=left,
x grid style={white!69.0196078431373!black},
xlabel={Number of Processors},
xmajorgrids,
xmin=0.794328234724281, xmax=125.892541179417,
xminorgrids,
xmode=log,
xtick style={color=black},
y grid style={white!69.0196078431373!black},
ylabel={Time (sec)},
ymajorgrids,
ymin=0.0874620350797613, ymax=19.4835836183402,
yminorgrids,
ymode=log,
ytick style={color=black},
ytick={0.001,0.01,0.1,1,10,100,1000},
yticklabels={\(\displaystyle {10^{-3}}\),\(\displaystyle {10^{-2}}\),\(\displaystyle {10^{-1}}\),\(\displaystyle {10^{0}}\),\(\displaystyle {10^{1}}\),\(\displaystyle {10^{2}}\),\(\displaystyle {10^{3}}\)}
]
\addplot [semithick, color0, mark=*, mark size=3, mark options={solid}]
table {%
1 3.742557806
4 1.358692766
16 0.4738858893
25 0.2759636748
64 0.1608038248
100 0.1118252813
};
\addlegendentry{Forward}
\addplot [semithick, color1, mark=*, mark size=3, mark options={solid}]
table {%
1 15.2387175252
4 5.0929829477
16 1.7030656633
25 1.0699440635
64 0.5695991209
100 0.460855982
};
\addlegendentry{Backward}
\end{axis}

\end{tikzpicture}}~
	\scalebox{0.6}{
\begin{tikzpicture}

\definecolor{color0}{rgb}{0.12156862745098,0.466666666666667,0.705882352941177}
\definecolor{color1}{rgb}{1,0.498039215686275,0.0549019607843137}

\begin{axis}[
legend cell align={left},
legend style={fill opacity=0.8, draw opacity=1, text opacity=1, at={(0.03,0.97)}, anchor=north west, draw=white!80!black},
log basis x={10},
tick align=outside,
tick pos=left,
x grid style={white!69.0196078431373!black},
xlabel={Number of Processors},
xmajorgrids,
xmin=0.794328234724281, xmax=125.892541179417,
xmode=log,
xtick style={color=black},
xtick={0.01,0.1,1,10,100,1000,10000},
xticklabels={\(\displaystyle {10^{-2}}\),\(\displaystyle {10^{-1}}\),\(\displaystyle {10^{0}}\),\(\displaystyle {10^{1}}\),\(\displaystyle {10^{2}}\),\(\displaystyle {10^{3}}\),\(\displaystyle {10^{4}}\)},
y grid style={white!69.0196078431373!black},
ylabel={Time (sec)},
ymajorgrids,
ymin=0.0, ymax=10,
ytick style={color=black}
]
\addplot [semithick, color0, mark=*, mark size=3, mark options={solid}]
table {%
1 1.0830292455
4 1.3877113241
9 1.4584997853
16 1.558316673
25 1.6177749694
36 1.658076819
49 1.6927434489
64 1.8140368681
81 1.8308102525
100 1.9432681085
};
\addlegendentry{Forward}
\addplot [semithick, color1, mark=*, mark size=3, mark options={solid}]
table {%
1 3.769420341
4 5.3936457647
9 6.0248365941
16 6.487088341
25 6.9462924425
36 7.1121010804
49 7.3269862706
64 7.4671808215
81 7.6033860796
100 7.8233449752
};
\addlegendentry{Backward}
\end{axis}

\end{tikzpicture}}
	\caption{Strong and weak scaling of acoustic wave equation.}
	\label{fig:acoustic}
\end{figure}

\begin{figure}[htpb]
	\centering 
	\scalebox{0.6}{
\begin{tikzpicture}

\definecolor{color0}{rgb}{0.12156862745098,0.466666666666667,0.705882352941177}
\definecolor{color1}{rgb}{1,0.498039215686275,0.0549019607843137}

\begin{groupplot}[group style={group size=2 by 1}]
\nextgroupplot[
legend cell align={left},
legend style={fill opacity=0.8, draw opacity=1, text opacity=1, at={(0.03,0.97)}, anchor=north west, draw=white!80!black},
log basis x={10},
log basis y={10},
tick align=outside,
tick pos=left,
title={Forward},
x grid style={white!69.0196078431373!black},
xlabel={Number of Processors},
xmajorgrids,
xmin=0.794328234724281, xmax=125.892541179417,
xminorgrids,
xmode=log,
xtick style={color=black},
y grid style={white!69.0196078431373!black},
ylabel={Time (sec)},
ymajorgrids,
ymin=0.265845012937478, ymax=42.1335951254345,
yminorgrids,
ymode=log,
ytick style={color=black},
ytick={0.01,0.1,1,10,100,1000},
yticklabels={\(\displaystyle {10^{-2}}\),\(\displaystyle {10^{-1}}\),\(\displaystyle {10^{0}}\),\(\displaystyle {10^{1}}\),\(\displaystyle {10^{2}}\),\(\displaystyle {10^{3}}\)}
]
\addplot [semithick, color0, mark=*, mark size=3, mark options={solid}]
table {%
1 1
4 2.75452839645133
16 7.89759283934011
25 13.561776957465
64 23.2740596229898
100 33.467904238574
};
\addlegendentry{Speedup}
\addplot [semithick, color1, mark=*, mark size=3, mark options={solid}]
table {%
1 1
4 0.688632099112832
16 0.493599552458757
25 0.542471078298599
64 0.363657181609215
100 0.33467904238574
};
\addlegendentry{Efficiency}

\nextgroupplot[
legend cell align={left},
legend style={fill opacity=0.8, draw opacity=1, text opacity=1, at={(0.03,0.97)}, anchor=north west, draw=white!80!black},
log basis x={10},
log basis y={10},
tick align=outside,
tick pos=left,
title={Backward},
x grid style={white!69.0196078431373!black},
xlabel={Number of Processors},
xmajorgrids,
xmin=0.794328234724281, xmax=125.892541179417,
xminorgrids,
xmode=log,
xtick style={color=black},
y grid style={white!69.0196078431373!black},
ylabel={Time (sec)},
ymajorgrids,
ymin=0.262653498360234, ymax=41.6277741527231,
yminorgrids,
ymode=log,
ytick style={color=black},
ytick={0.01,0.1,1,10,100,1000},
yticklabels={\(\displaystyle {10^{-2}}\),\(\displaystyle {10^{-1}}\),\(\displaystyle {10^{0}}\),\(\displaystyle {10^{1}}\),\(\displaystyle {10^{2}}\),\(\displaystyle {10^{3}}\)}
]

{
\hspace*{1cm}
\addplot [semithick, color0, mark=*, mark size=3, mark options={solid}]
table {%
1 1
4 2.99210063761981
16 8.94781560898375
25 14.2425366381782
64 26.7534077319535
100 33.0661163582336
};
\addlegendentry{Speedup}
\addplot [semithick, color1, mark=*, mark size=3, mark options={solid}]
table {%
1 1
4 0.748025159404953
16 0.559238475561484
25 0.569701465527128
64 0.418021995811774
100 0.330661163582336
};
\addlegendentry{Efficiency}
}
\end{groupplot}

\end{tikzpicture}}
	\caption{Speedup and efficiency for acoustic wave equations.}
	\label{fig:acoustic_speedup_and_efficiency}
\end{figure}
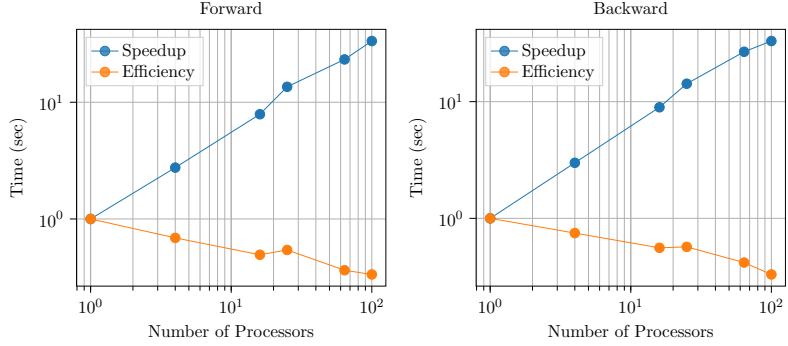

We also show the speedup and efficiency for the strong scaling case in \Cref{fig:acoustic_speedup_and_efficiency}. We can achieve more than 20 times acceleration by using 100 processors (3200 cores in total) and the trend is not slowing down at this scale.

\subsection{Elastic Wave Equation}

\begin{figure}[htpb]
	\centering 
	\includegraphics[width=0.18\textwidth]{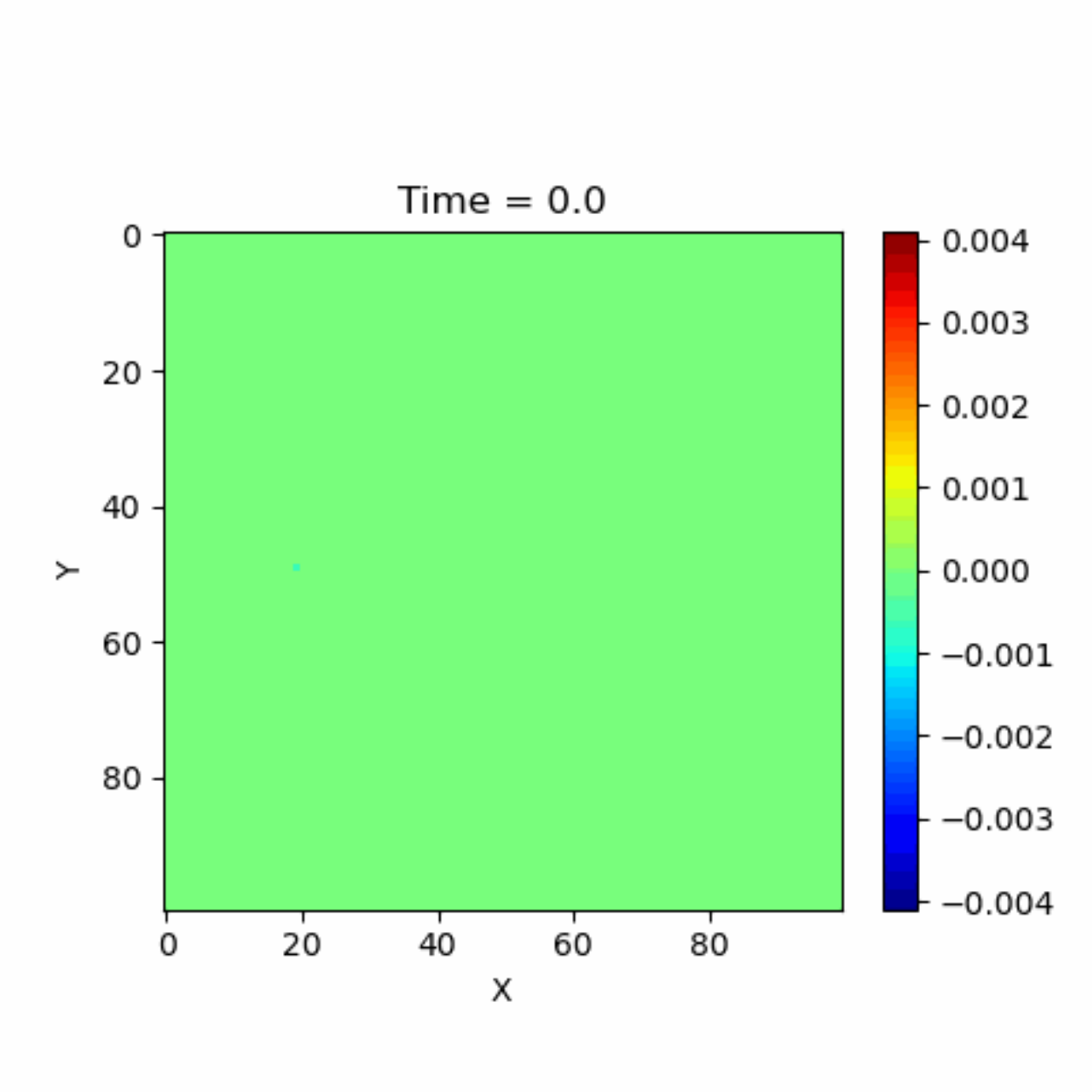}~
	\includegraphics[width=0.18\textwidth]{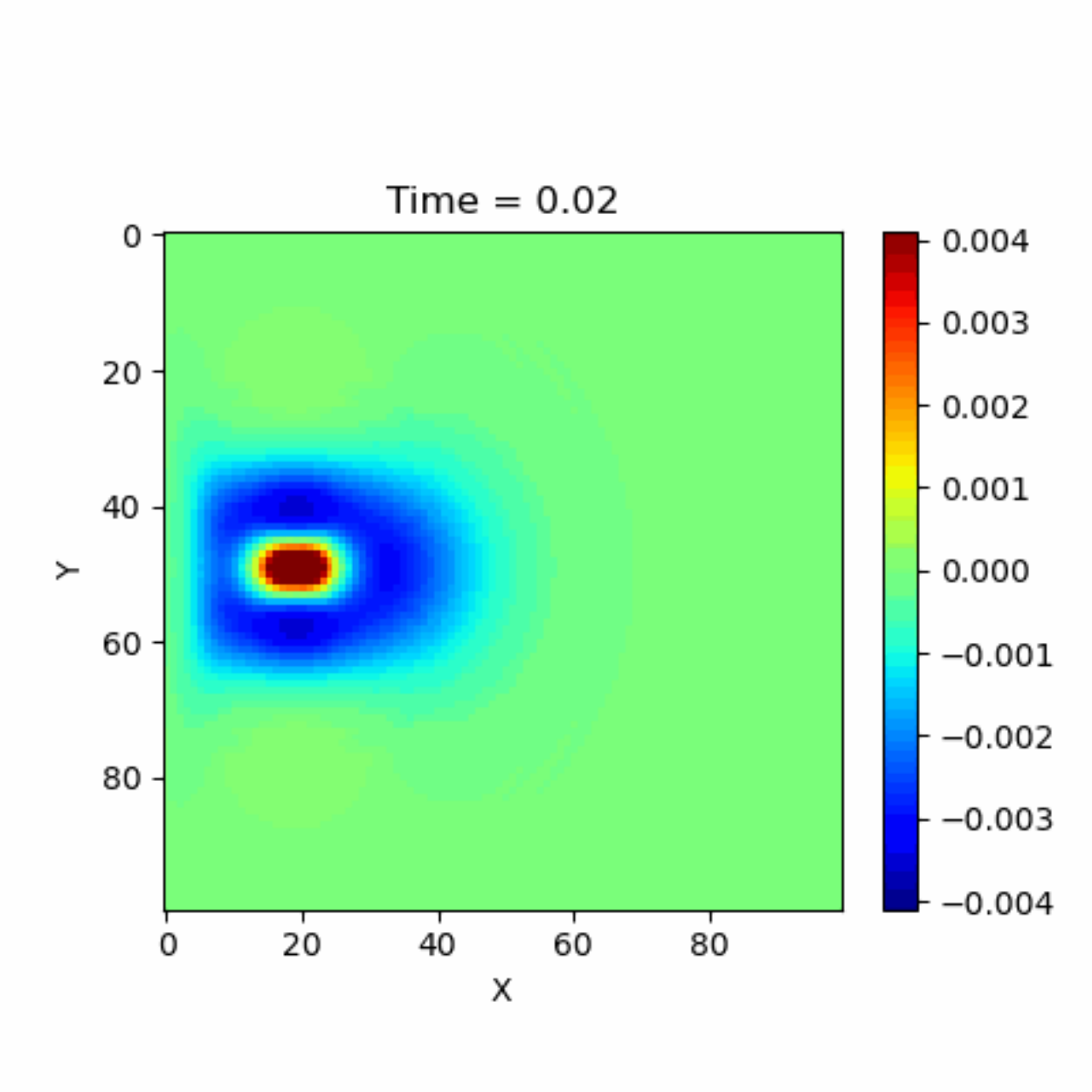}~
	\includegraphics[width=0.18\textwidth]{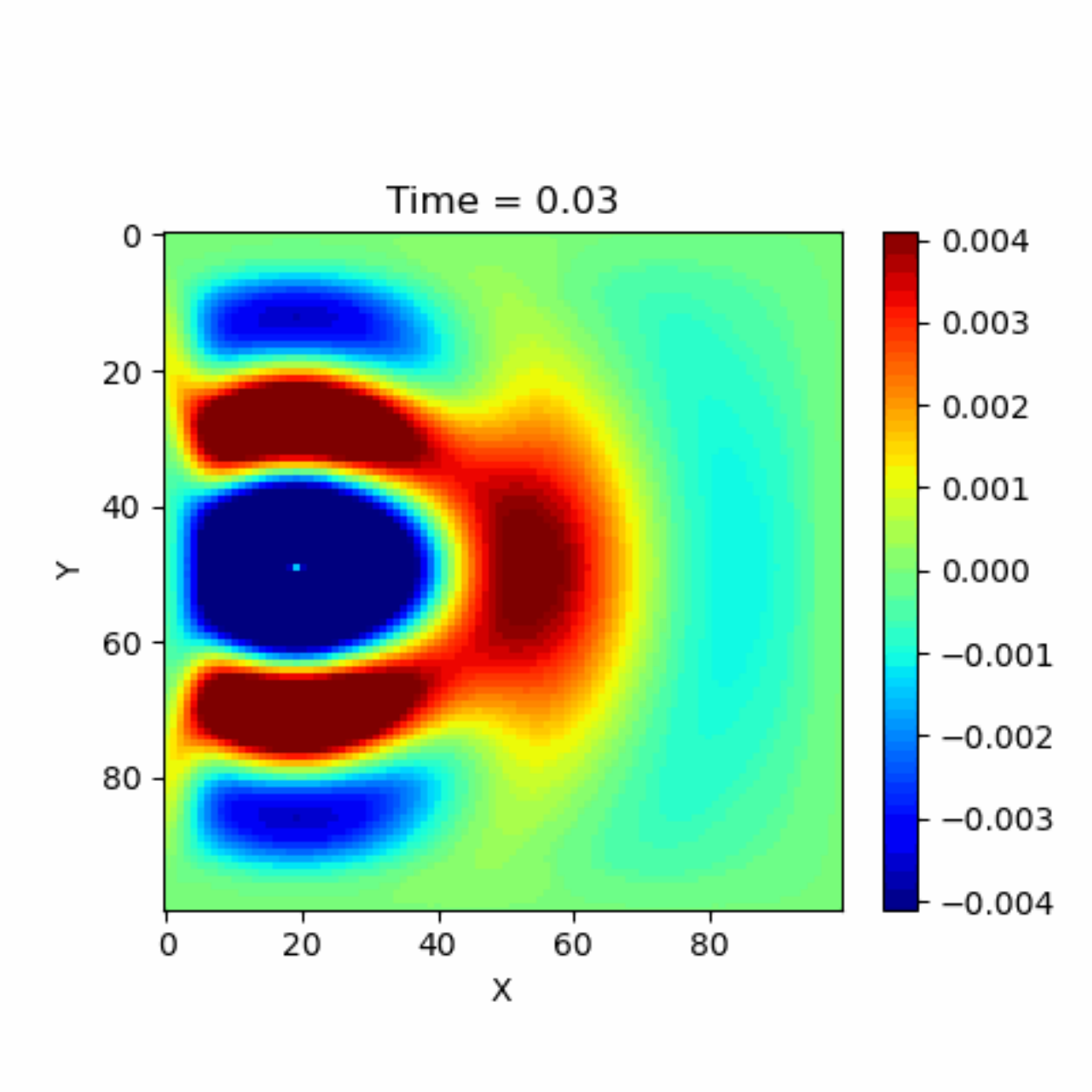}~
	\includegraphics[width=0.18\textwidth]{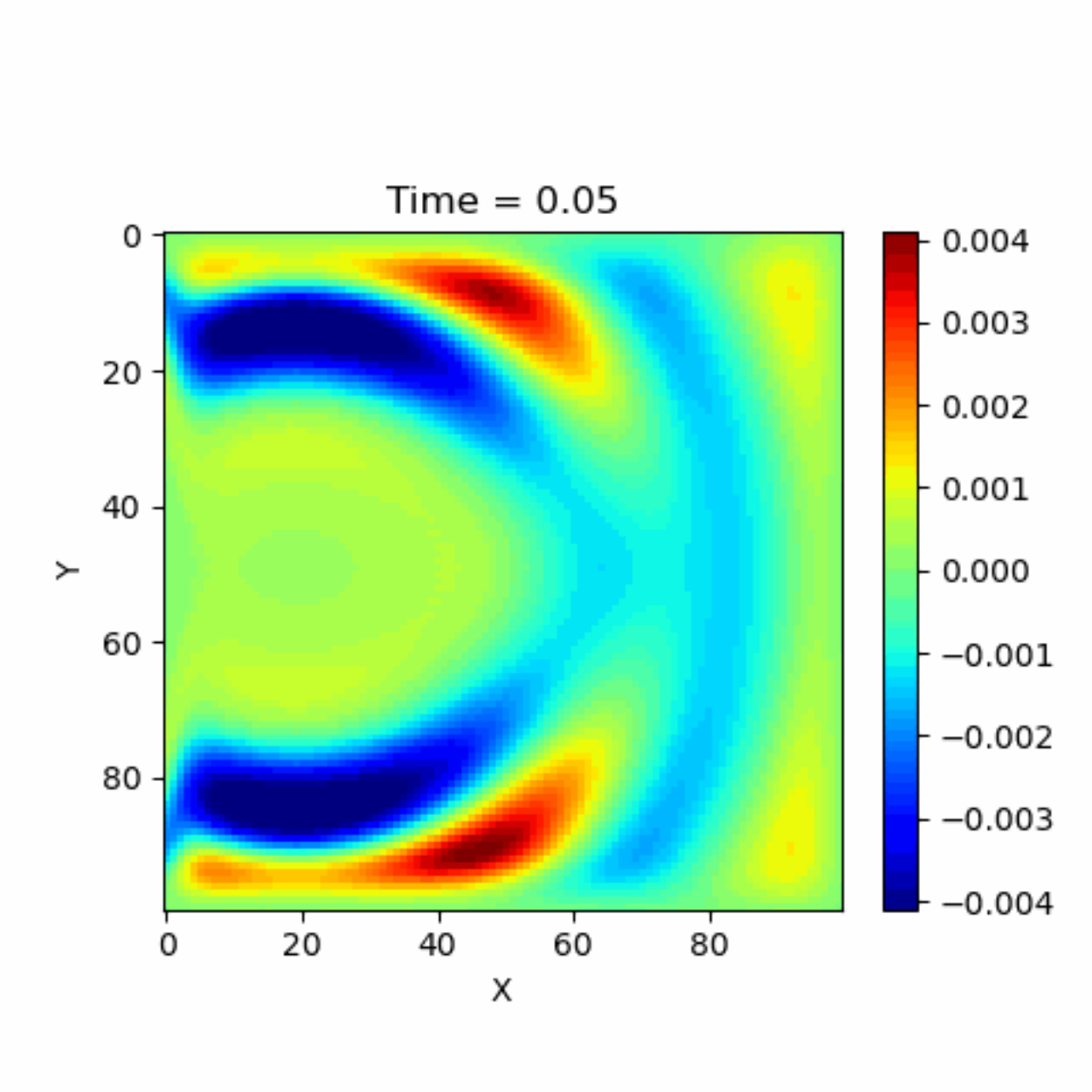}~
	\includegraphics[width=0.18\textwidth]{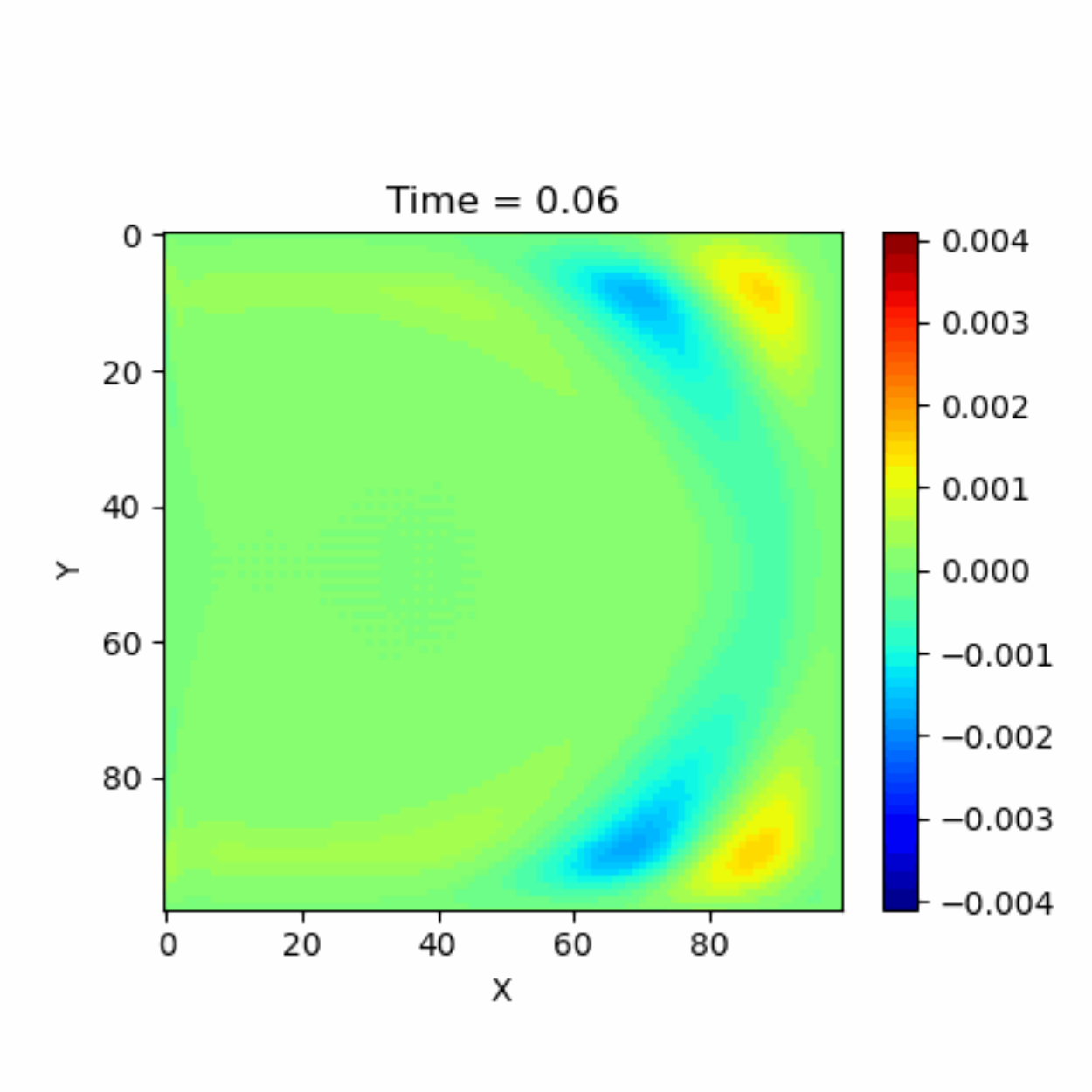}
	\caption{Snapshots of $v_1$ for the elastic wave propagation.}
	\label{fig:elastic_snapshots}
\end{figure}

In the last example, we consider the elastic wave equation \cite{li2020coupled}
$$\begin{aligned} \rho \frac{\partial v_i}{\partial t} &= \sigma_{ij,j} + \rho f_i \\ \frac{\partial \sigma_{ij}}{\partial t} &= \lambda v_{k, k} + \mu (v_{i,j}+v_{j,i}) \end{aligned}$$
where $v$ is the velocity, $\sigma$ is the stress tensor, $\rho$ is the density, and $\lambda$ and $\mu$ are the Lam\'e constants. Similar to the acoustic equation, we use the PML boundary conditions and have observations on the surface. However, the inversion parameters are now spatially varying $\rho$, $\lambda$ and $\mu$. \Cref{fig:elastic_snapshots} shows snapshots of the wave propagation.

As an example, we approximate $\lambda$ by a deep neural network
$$\lambda(x,y) = \mathcal{N}_\theta(x,y)$$
and the other two parameters are kept fixed.

We use the same geometry settings as the acoustic wave equation case. Note that elastic wave equation has more state variables as well as auxilliary fields, and thus is more memory demanding. The huge memory cost calls for a distributed framework, especially for large-scale problems.

Additionally, we use fourth-order finite difference scheme for discretizing Equation 2. This scheme requires us to exchange two layers on the boundaries for each block in the mesh. This data communication is implemented using MPI, i.e., \texttt{mpi\_halo\_exchange2}.

\begin{figure}[htpb]
	\centering 
	\scalebox{0.6}{
\begin{tikzpicture}

\definecolor{color0}{rgb}{0.12156862745098,0.466666666666667,0.705882352941177}
\definecolor{color1}{rgb}{1,0.647058823529412,0}

\begin{axis}[
legend cell align={left},
legend style={fill opacity=0.8, draw opacity=1, text opacity=1, draw=white!80!black},
log basis x={10},
log basis y={10},
tick align=outside,
tick pos=left,
x grid style={white!69.0196078431373!black},
xlabel={Number of Processors},
xmajorgrids,
xmin=0.794328234724281, xmax=125.892541179417,
xminorgrids,
xmode=log,
xtick style={color=black},
y grid style={white!69.0196078431373!black},
ylabel={Time (sec)},
ymajorgrids,
ymin=0.369407138088942, ymax=50.4654841733428,
yminorgrids,
ymode=log,
ytick style={color=black},
ytick={0.01,0.1,1,10,100,1000},
yticklabels={\(\displaystyle {10^{-2}}\),\(\displaystyle {10^{-1}}\),\(\displaystyle {10^{0}}\),\(\displaystyle {10^{1}}\),\(\displaystyle {10^{2}}\),\(\displaystyle {10^{3}}\)}
]
\addplot [semithick, color0, mark=*, mark size=3, mark options={solid}]
table {%
1 15.2674026251
4 6.6159459011
16 1.7839082127
25 1.0301020655
64 0.5982715645
100 0.4619266005
};
\addlegendentry{Forward}
\addplot [semithick, color1, mark=*, mark size=3, mark options={solid}]
table {%
1 40.3577322903
4 17.7924579277
16 4.7683545434
25 2.9973443985
64 1.6802775259
100 1.3808860655
};
\addlegendentry{Backward}
\end{axis}

\end{tikzpicture}}~
	\scalebox{0.6}{
\begin{tikzpicture}

\definecolor{color0}{rgb}{0.12156862745098,0.466666666666667,0.705882352941177}
\definecolor{color1}{rgb}{1,0.498039215686275,0.0549019607843137}

\begin{axis}[
legend cell align={left},
legend style={fill opacity=0.8, draw opacity=1, text opacity=1, at={(0.91,0.5)}, anchor=east, draw=white!80!black},
log basis x={10},
tick align=outside,
tick pos=left,
x grid style={white!69.0196078431373!black},
xlabel={Number of Processors},
xmajorgrids,
xmin=0.794328234724281, xmax=125.892541179417,
xmode=log,
xtick style={color=black},
xtick={0.01,0.1,1,10,100,1000,10000},
xticklabels={\(\displaystyle {10^{-2}}\),\(\displaystyle {10^{-1}}\),\(\displaystyle {10^{0}}\),\(\displaystyle {10^{1}}\),\(\displaystyle {10^{2}}\),\(\displaystyle {10^{3}}\),\(\displaystyle {10^{4}}\)},
y grid style={white!69.0196078431373!black},
ylabel={Time (sec)},
ymajorgrids,
ymin=0, ymax=50,
ytick style={color=black}
]
\addplot [semithick, color0, mark=*, mark size=3, mark options={solid}]
table {%
1 3.6454164948
4 5.4364835492
9 5.7843076207
16 6.0657446765
25 6.4206062678
36 6.7304453538
49 6.9285739692
64 7.0808431315
81 7.3740807225
100 7.4873187403
};
\addlegendentry{Forward}
\addplot [semithick, color1, mark=*, mark size=3, mark options={solid}]
table {%
1 19.7673796185
4 31.2877235353
9 32.6022873706
16 35.4975575161
25 35.8438810998
36 39.1725659084
49 36.9233598893
64 38.8343191014
81 38.5663559224
100 39.9757250535
};
\addlegendentry{Backward}
\end{axis}

\end{tikzpicture}}
	\caption{Strong and weak scaling of elastic wave equation.}
	\label{fig:elastic}
\end{figure}
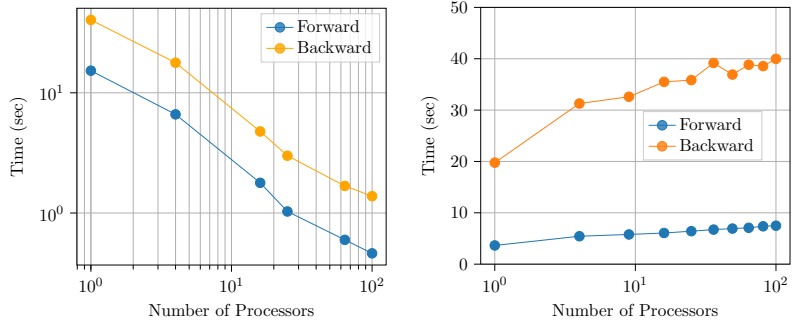

\Cref{fig:elastic} shows the strong and weak scaling. Again, we see that the weak scaling of the implementation is quite effective because the runtime increases mildly even if we increase the number of processors from 1 to 100.

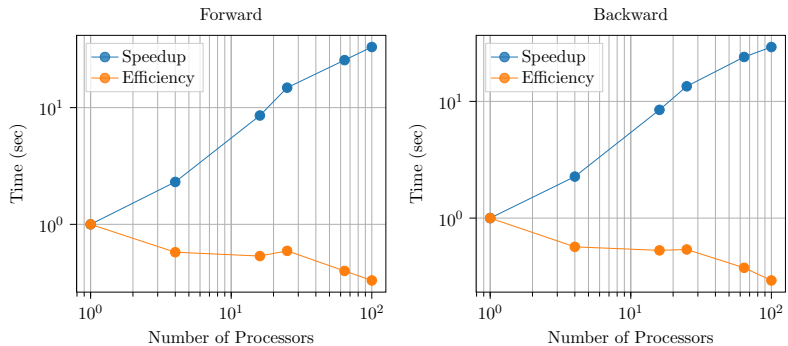
\begin{figure}[htpb]
	\centering 
	\scalebox{0.6}{
\begin{tikzpicture}

\definecolor{color0}{rgb}{0.12156862745098,0.466666666666667,0.705882352941177}
\definecolor{color1}{rgb}{1,0.498039215686275,0.0549019607843137}

\begin{groupplot}[group style={group size=2 by 1}]
\nextgroupplot[
legend cell align={left},
legend style={fill opacity=0.8, draw opacity=1, text opacity=1, at={(0.03,0.97)}, anchor=north west, draw=white!80!black},
log basis x={10},
log basis y={10},
tick align=outside,
tick pos=left,
title={Forward},
x grid style={white!69.0196078431373!black},
xlabel={Number of Processors},
xmajorgrids,
xmin=0.794328234724281, xmax=125.892541179417,
xminorgrids,
xmode=log,
xtick style={color=black},
y grid style={white!69.0196078431373!black},
ylabel={Time (sec)},
ymajorgrids,
ymin=0.262538008482162, ymax=41.6094702405677,
yminorgrids,
ymode=log,
ytick style={color=black},
ytick={0.01,0.1,1,10,100,1000},
yticklabels={\(\displaystyle {10^{-2}}\),\(\displaystyle {10^{-1}}\),\(\displaystyle {10^{0}}\),\(\displaystyle {10^{1}}\),\(\displaystyle {10^{2}}\),\(\displaystyle {10^{3}}\)}
]
\addplot [semithick, color0, mark=*, mark size=3, mark options={solid}]
table {%
1 1
4 2.3076673922865
16 8.55840144487721
25 14.8212523170598
64 25.5191848167806
100 33.0515770440027
};
\addlegendentry{Speedup}
\addplot [semithick, color1, mark=*, mark size=3, mark options={solid}]
table {%
1 1
4 0.576916848071625
16 0.534900090304826
25 0.592850092682394
64 0.398737262762198
100 0.330515770440027
};
\addlegendentry{Efficiency}

\nextgroupplot[
legend cell align={left},
legend style={fill opacity=0.8, draw opacity=1, text opacity=1, at={(0.03,0.97)}, anchor=north west, draw=white!80!black},
log basis x={10},
log basis y={10},
tick align=outside,
tick pos=left,
title={Backward},
x grid style={white!69.0196078431373!black},
xlabel={Number of Processors},
xmajorgrids,
xmin=0.794328234724281, xmax=125.892541179417,
xminorgrids,
xmode=log,
xtick style={color=black},
y grid style={white!69.0196078431373!black},
ylabel={Time (sec)},
ymajorgrids,
ymin=0.23215011758426, ymax=36.7933140988341,
yminorgrids,
ymode=log,
ytick style={color=black},
ytick={0.01,0.1,1,10,100,1000},
yticklabels={\(\displaystyle {10^{-2}}\),\(\displaystyle {10^{-1}}\),\(\displaystyle {10^{0}}\),\(\displaystyle {10^{1}}\),\(\displaystyle {10^{2}}\),\(\displaystyle {10^{3}}\)}
]
{
\hspace*{1cm}
\addplot [semithick, color0, mark=*, mark size=3, mark options={solid}]
table {%
1 1
4 2.2682494152463
16 8.46366014166462
25 13.4644962088763
64 24.0184919861279
100 29.2259682377829
};
\addlegendentry{Speedup}
\addplot [semithick, color1, mark=*, mark size=3, mark options={solid}]
table {%
1 1
4 0.567062353811576
16 0.528978758854039
25 0.538579848355054
64 0.375288937283249
100 0.292259682377829
};
\addlegendentry{Efficiency}
}
\end{groupplot}

\end{tikzpicture}}
	\caption{Speedup and efficiency for elastic wave equations.}
	\label{fig:elastic_speedup_and_efficiency}
\end{figure}

\Cref{fig:elastic_speedup_and_efficiency} shows the speedup and efficiency for the strong scaling. We can achieve more than 20 times speedup when using 100 processors.

\section{Conclusion}

In this paper we proposed a framework for distributed optimization in scientific machine learning for solving inverse problems. The key is to express numerical simulation using a computational graph. The data communication operations are treated as nodes in the comptuational graph. This view makes the implementation of inverse modeling algorithms quite flexible (modular and testable) and conceptually simple. We also proposed a method to convert existing gradient-based optimization algorithms to MPI-enabled optimizers. These ideas are implemented in the ADCME library, which provides a set of MPI primitives, which can back-propagate gradients. We can also develop customized data communication operators to tailor to specific applications for better performance. To demonstrate the effectiveness of the proposed algorithm, we have trained a neural network that is coupled with either a Poisson's equation or a wave equation. The PDE is solved numerically and in a distributed way. The results show that our method provides a promising approach towards scalable inverse modeling.

\section*{\refname}

\bibliographystyle{unsrt}
\bibliography{ref}
\end{document}